\documentclass[prd,twocolumn,superscriptaddress,floatfix,amsmath,amssymb,amsfonts]{revtex4-1}

\usepackage{float} 

\usepackage{comment}
\usepackage[normalem]{ulem}
\usepackage[english]{babel}
\usepackage{graphicx}
\usepackage{dcolumn}
\usepackage{bm}
\usepackage{blindtext}
\usepackage{verbatim}
\usepackage{mathrsfs}
\usepackage{musicography}
\usepackage{amsmath}
\usepackage{cancel}
\usepackage{physics}
\usepackage{epstopdf}
\usepackage{mathtools}
\usepackage{color}
\usepackage[usenames,dvipsnames]{pstricks}
\usepackage{epsfig}
\usepackage{pst-grad} 
\usepackage{pst-plot} 
\usepackage{hyperref}
\usepackage{verbatim}

\newcommand{\mf}{\mathsf}

\newcommand{\ii}{\mathrm{i}}

\newcommand{\barsf}[1]{\bar{\mathsf{#1}}}
\newcommand{\timelike}{\musNatural}
\newcommand{\lightlike}{\rotatebox[origin=c]{135}{\musNatural}}
\newcommand{\spacelike}{\rotatebox[origin=c]{90}{\musNatural}}

\begin{document}

\title{Broken covariance of particle detector models in relativistic quantum information}

\author{Eduardo Mart\'{i}n-Mart\'{i}nez}
\email{emartinmartinez@uwaterloo.ca}
\affiliation{Institute for Quantum Computing, University of Waterloo, Waterloo, Ontario, N2L 3G1, Canada}
\affiliation{Department of Applied Mathematics, University of Waterloo, Waterloo, Ontario, N2L 3G1, Canada}
\affiliation{Perimeter Institute for Theoretical Physics, Waterloo, Ontario, N2L 2Y5, Canada}

\author{T. Rick Perche}
\email{trickperche@perimeterinstitute.ca}
\affiliation{Perimeter Institute for Theoretical Physics, Waterloo, Ontario, N2L 2Y5, Canada}

\affiliation{Instituto de F\'{i}sica Te\'{o}rica, Universidade Estadual Paulista, S\~{a}o Paulo, S\~{a}o Paulo, 01140-070, Brazil}

\author{Bruno de S. L. Torres}
\email{bdesouzaleaotorres@perimeterinstitute.ca}
\affiliation{Perimeter Institute for Theoretical Physics, Waterloo, Ontario, N2L 2Y5, Canada}

\affiliation{Instituto de F\'{i}sica Te\'{o}rica, Universidade Estadual Paulista, S\~{a}o Paulo, S\~{a}o Paulo, 01140-070, Brazil}

\begin{abstract}

We show that the predictions of \textcolor{black}{commonly used} spatially smeared particle detectors coupled to quantum fields are not generally covariant outside the pointlike limit. This lack of covariance manifests itself as an ambiguity in the time-ordering operation. We analyze how the breakdown of covariance affects typical detector models in quantum field theory such as the UDW model. Specifically, we show how the violations of covariance depend on the state of the detectors-field system, the shape and state of motion of the detectors, and the spacetime geometry. Furthermore, we provide the tools to explicitly evaluate the magnitude of the violation, and identify the regimes where the predictions of smeared detectors are either exactly or approximately covariant in perturbative analyses, \textcolor{black}{thus providing limits of validity of smeared particle detector models}.



\end{abstract}

\maketitle

\section{Introduction}    

    Particle detector models~\cite{Unruh1976,Unruh-Wald,DeWitt} have become an ubiquitous concept in the study of fundamental problems in quantum field theory (QFT). They provide a way of circumventing some of the conceptual and technical issues associated with the notion of measurement of localized field observables \cite{Sorkin,Benincasa_2014,fewster3,borsten}, and also yield an operationally appealing approach to common  phenomenology in QFT in curved spacetimes such as the Unruh and Hawking effects (see, e.g., \cite{Unruh1976,Sciama1977,Unruh-Wald,Takagi,Louko}). Beyond their value as a fundamental tool, particle detector models are commonly employed in concrete setups in relativistic quantum information and quantum optics to model the light-matter interaction in relativistic regimes (see, e.g. \cite{Pozas2016,eduardo}).
    
    Common desired features of particle detector models include being localized, controllable and measurable nonrelativistic quantum systems that couple to a quantum field in a finite region of spacetime. Historically~\cite{DeWitt}, particle detectors have been typically considered to be pointlike objects which interact with a quantum field along timelike curves representing their trajectories. There are, however, good reasons to extend the model beyond pointlike detectors, thus including some spatial extension to the system. One reason is to regularize UV divergences in the predictions of the theory by introducing a finite lenghtscale for the size of the detector~\cite{Schlicht,Jorma}. Smeared detectors are also more appealing from the point of view of algebraic quantum field theory, where field observables are directly linked to field operators that are smeared in both time and space~\cite{kasia}, and to which it is natural to couple our detectors. Finally, one could also argue for the need for smeared particle detector models due to the fact that in all physically realistic scenarios, the devices being used as detector---for instance, an atom coupling to the electromagnetic field~\cite{ScullyBook,Pozas2016,eduardo}---is not a pointlike object, but has in fact some nontrivial spatial extension.

    
    Smeared particle detectors, however, are not devoid of their own issues. In particular, coupling a single nonrelativistic degree of freedom of the detector to a region of spacetime with finite spatial extension implies ``faster than light'' coupling of the internal constituents of the detector. In other words, one single detector's degree of freedom ``feels'' the interaction with the field simultaneously at spacelike separated points. \textcolor{black}{ This does not mean that smeared detectors should be avoided in a relativistic description of measurement in QFT. Rather, this seems intuitively compatible with the assumption that the detector is a non-relativistic system.  Indeed, particle detector models do not intend to yield a fundamental description of reality, but rather, to provide an approximate description of measurements in quantum fields that is valid under certain regimes. The effects that the detector's `non-locality' of the coupling may have} on the causal behaviour of the detector model were analyzed in~\cite{martin-martinez2015}, where it was shown that as long as predictions are taken at times longer than the light-crossing time of the detectors' lengthscales, smeared particle detector models cannot signal faster than light. Furthermore, following on this, in recent work~\cite{us}, it was discussed that there are ways to covariantly prescribe the coupling between smeared detectors and fields. However, even when the detector-field Hamiltonian density is covariantly prescribed, one may wonder whether there may still be issues with the covariance of the time evolution generated by this Hamiltonian density due to the non-local nature of the coupling of smeared detectors.  


    Indeed, taking the common Hamiltonian formulation for particle detector physics, we can ask how these non-locality issues affect the time evolution operator given by the time-ordered exponential of the Hamiltonian for the system. Although in nonrelativistic physics time is an external, absolute parameter, when considering relativistic scenarios, one has to address the issues and subtleties that arise from different choices of a time parameter. In particular, each observer has their own rest spaces and proper times, and therefore the notion of time order may become frame dependent. Namely, the ordering of events in spacetime according to different time coordinates will only be unambiguous if the events are timelike or null separated. If, on the other hand, two events are spacelike separated, one can find observers that see either event happening before or after the other. 
    
    In the case of particle detectors, first principle arguments tell us that it is physically justified  to prescribe the interaction in the reference frame of the detector's center of mass~\cite{eduardo,us}. However, if the interaction between the field and the detector is spatially smeared, there will be spacelike separated events in the `worldtube' of the detector. This means that the ambiguity in time ordering will impact smeared detector setups, since certainly time-ordering with respect to the detector's centre of mass proper time will in general not be equivalent to time-ordering with respect to a different frame. If taken at face value, this would be catastrophic for a detector model of a quantum field theory: suddenly, time evolution and all its predictions would be reference frame dependent.
    
    General covariance is an important foundational point of modern theoretical physics: fundamental theories must be independent of the (strictly mathematical) choice of the coordinates used to describe the laws of physics. Even though the detector based approach for probing quantum fields is not intended to be a fundamental description of nature, it is still important that its predictions are generally covariant if we are to give them physical meaning in terms of features of the quantum field. Moreover, particle detectors are used in scenarios where covariance plays an important role, such as entanglement harvesting (see, e.g.,~\cite{Valentini1991,Reznik1,reznik2,Retzker2005,RalphOlson1,RalphOlson2,Nick,Cosmo,Salton:2014jaa,Pozas-Kerstjens:2015}), where multiple detectors are present and the causal relations between the interactions of the many detectors are relevant.
    
    

    In the present paper, we study in detail how the spatial smearing of an UDW detector breaks covariance. First, we show that all predictions made for a system of pointlike detectors with covariant Hamiltonian densities (prescribed as in \cite{us}) are coordinate independent. In other words, systems of many pointlike particle detectors in general spacetime backgrounds are fully covariant. 
    
    We then explicitly analyze the time evolution operator for smeared detectors and calculate (up to lowest nontrivial order) the magnitude of the violation of covariance due to the detectors' finite size. In particular, we show how predictions made in different coordinate systems with different notions of time-ordering deviate from each other as a function of the field-detector state, the size and shape of the detector, as well as the geometry of spacetime. We will show that if the detector is initially in a statistical mixture of states of well defined energy (eigenstates of the free Hamiltonian, thermal states, etc), then the violations of covariance are of third order (and in many cases fourth order) in the coupling strength between the detector and field. This means that predictions associated to different choices of time parameters are equivalent at the order in perturbation theory where many important phenomena manifest (e.g., entanglement harvesting, detection of the Unruh effect, etc.). Furthermore, for the cases where the violations of covariance are of leading order, we discuss in what regimes they can be made negligible. Namely, approximate covariance is restored when several requirements are met: 1) the relative motion of the detectors with respect to the frame in which we are computing should not be extreme; 2) the curvature around the detectors should also be small enough; and 3) the predictions are only taken for times much longer than the light-crossing time of each of the detectors in their respective proper frames, as well as in the coordinate frame we use to calculate.



\section{Review of Spacetime intervals in curved spacetimes}\label{intervals}
    
   For the purposes of this work, it is convenient to review the notions of timelike, null and spacelike separation in curved spacetimes. In Minkowski spacetime it is easy to define the notion of spacelike and timelike separation of two events $\mathsf{p}$ and $\mathsf{q}$. If we let $\Delta \mathsf{x} = \mathsf{p}-\mathsf{q}$, we say that the two events are spacelike separated in the case in which $\eta_{\mu\nu}\Delta \mathsf{x}^\mu\Delta \mathsf{x}^\nu > 0$ and that they are timelike separated if $\eta_{\mu\nu}\Delta \mathsf{x}^\mu\Delta \mathsf{x}^\nu < 0$, where $\eta_{\mu\nu}$ stands for the metric in inertial coordinates. If $\mathsf{p}$ and $\mathsf{q}$ are spacelike separated in Minkowski spacetime, it is always possible to find an inertial timelike observer that sees both events simultaneously. If they are timelike separated, there is always an inertial timelike trajectory that goes through both of the events. Also, we say that two events are null separated if the norm of $\Delta \mf x$ is zero. Null separated events are connected by a ray of light. 
    
    These concepts provide very useful insight about the causal structure of Minkowski spacetime, in the sense that events that are spacelike separated have no causal influence over one another. In the context of quantum field theory, this fact manifests itself as the microcausality condition: the commutator of quantum fields in spacelike separated regions vanishes. In Minkowski spacetimes, we can simply say timelike separated events are those events that can be connected by timelike curves, and the analogous holds for null separated events. The points that are spacelike separated are the ones that do not fit any of the categories before. 
   
   In curved spacetimes, however, defining global notions analogous to timelike, null and spacelike separations is  more delicate \cite{Wald1}. First, we assume that we have a spacetime $\mathcal{M}$ with a metric $\mathsf{g}$ that is globally hyperbolic and time orientable. Given a point $\mf{p}$, we then define the set of chronological events related to $\mathsf{p}$  as the set of all points that can be connected to $\mathsf{p}$ by a timelike curve. We denote the set of chronological events related to $\mathsf{p}$ by $I(\mathsf{p})$. This can be shown to be an open set \cite{Wald1} and it corresponds to the interior of the lightcone in Minkowski spacetime. 
    
    We then define the set of null separated events $N(\mf p) = \overline{I(\mathsf{p})}\setminus I(\mathsf{p})$ as the boundary of the closure of $I(\mathsf{p})$. In Minkowski  $N(\mf p)$ corresponds to the set of points that are in the boundary of the lightcone of $\mf p$. It should be noted however that in general this set might contain points that are not causally connected to $\mathsf{p}$ (See again \cite{Wald1} for an example). However, all null curves that go through $\mathsf{p}$ are contained in $N(\mf p)$. Note that as $N(\mf p)$ is the boundary of a region, it possesses one dimension less than the spacetime it is contained in, and volume integrals performed over it yield zero.
    
    We then define the set of non-chronological events related to $\mathsf{p}$ as $S(\mathsf{p}) = \mathcal{M}\setminus \overline{I(\mathsf{p})}$. In Minkowski this is equivalent to the region outside the lightcone of $\mathsf{p}$. This set is always open since it is the complement of a closed set, and no event in $S(\mathsf{p})$ is causally connected to the point $\mathsf{p}$.
    
    Having generalizations of spacelike and timelike separation, we notice that $\mathsf{q}\in I(\mathsf{p}) \Leftrightarrow \mathsf{p}\in I(\mathsf{q})$, so that we can define the relation between events  `belonging to the chronological set of one another' that we notate $\mathsf{p}\timelike \mathsf{q}$. This relation is what we define as timelike separation. Analogously, $\mathsf{q}\in N(\mathsf{p}) \Leftrightarrow \mathsf{p}\in N(\mathsf{q})$ so that we can define the  relation between events `belonging to the null set of one another' that we denote by $\mf p \lightlike \mf q$. This is what we will call null separation here, although it should be noted that not all null separated events can be connected by a lightlike curve \cite{Wald1}. 
    
    \color{black}
     Consider two future-oriented timelike differential forms $\dd t$ and $\dd t'$ which foliate spacetime by achronal surfaces, and are associated to two coordinate systems
    \color{black}
    $R\equiv(t,\bm x)$ and $R'\equiv(t',\bm x')$. Now, take two events $\mathsf{p}$ and $\mathsf{q}$ of  coordinates $(p^0,p^i), (p^{0'},p^{i'})$ and $(q^0,q^i), (q^{0'},q^{i'})$ respectively in $R$ and $R'$. If $\mf p \timelike \mf q$ or $\mf p \lightlike \mf q$
    , we have that the sign of $p^0-q^0$ and $p^{0'}-q^{0'}$ is the same. Also, we will only have $p^0-q^0$ (or $p^{0'}-q^{0'}$) equal zero if $\mf p=\mf q$. Therefore, the notion of time ordering for these events is unambiguous and coordinate independent (hence reference frame independent), \textcolor{black}{provided that the surfaces of constant $t$ and $t'$ are achronal}. This will be particularly useful for the discussion of the meaning of the time ordering operation in quantum mechanics in curved spacetimes.
    
    We also define the relation $\mathsf{p}\spacelike \mathsf{q}$ in the case in which  $\mathsf{q} \in S(\mathsf{p})\Leftrightarrow\mathsf{p}\in S(\mathsf{q})$. Relevant to this paper, notice that since in this case the points $\mathsf{p}$ and $\mathsf{q}$ are not causally connected, the microcausality condition imposes that the commutator of a scalar quantum field evaluated at them must vanish. 
    
    \section{The Unruh-DeWitt Model in Curved Spacetimes}\label{UDWmodel}
    

    
    
    
    
    
    
    To model the interaction of a particle detector and a quantum field in curved spacetimes we use a smeared Unruh-DeWitt (UDW) detector~\cite{DeWitt,Unruh-Wald}. That is, a two-level system interacting with a free scalar field through a minimally coupled (for simplicity) action. The UDW model captures most of the fundamental features of the light-matter interaction (barring the exchange of angular momentum~\cite{EduardoOld,eduardo}) and hence one could think of this detector as modelling the interaction of atomic probes and the electromagnetic field~\cite{Pozas2016,eduardo}. \color{black}
    
    To describe the quantum field and detector, we assume that we have a globally hyperbolic $D = n+1$ dimensional spacetime $\mathcal{M}$. Under these assumptions, the action for a (minimally coupled) classical real scalar field can be written as
    \begin{equation}\label{actionField}
        S[\phi] = \int \!\dd\mathcal{V}\left(-\dfrac{1}{2}\nabla_{\mu}\phi\nabla^{\mu}\phi  - \dfrac{1}{2}m^2 \phi ^2\right),
    \end{equation}
    where $\dd \mathcal{V}$ is the invariant volume element of spacetime, given by
    \begin{equation}
        \dd \mathcal{V} \equiv \sqrt{- g} \dd^D \mathsf{x} = \sqrt{- \bar{g}} \dd^D \bar{\mathsf{x}}.
    \end{equation}    
    When extremized, the action in Eq. \eqref{actionField} yields the  Klein-Gordon equation of motion for the field $\phi$,
    \begin{equation}\label{KG}
        \nabla_\mu \nabla^\mu \phi - m^2 \phi = 0.
    \end{equation}
     At this point we can pick a complete set of solutions to Eq. \eqref{KG}, $\{u_{\bm k}(\mf x)\}$ which is orthonormal with respect to the Klein-Gordon inner product~\cite{Takagi,Wald2}. That is,
    \begin{align}
        (u_{\bm k},\bm u_{\bm k'}) &= \delta(\bm k - \bm k'),\\
        (u_{\bm k}^*,\bm u_{\bm k'}^*) &= -\delta(\bm k - \bm k'),\\
        (u_{\bm k},\bm u_{\bm k'}^*) &=0.
    \end{align}
    This allows one to write any classical solution $\phi(\mf x)$ as a linear combination of the $u_{\bm k}(\mf x)$ and $u_{\bm k}^*(\mf x)$:
    \begin{equation}\label{modeExp}
        \phi(\mf x) = \int \dd^{n} \!\bm{k}\left({a}^*_{\bm{k}} u_{\bm{k}}^*(\mathsf x)+{a}_{\bm{k}} u_{\bm{k}}( \mathsf x) \right),
    \end{equation}
    where ${a_{\bm k} = (u_{\bm k},\phi)}$. Notice that the field $\phi(\mf x)$ can be completely determined from the coefficients $a_{\bm k}$, which can be calculated in any Cauchy surface $\Sigma$ provided that both $\phi$ and its normal derivative to the surface are specified in $\Sigma$. This procedure is independent of the mode expansion performed and of the Cauchy surface chosen to prescribe the initial conditions.
    

   To canonically quantize the field $\phi(\mf x)$, one must first define the conjugate momentum to the field, $\pi(\mf x)$. The form of $\pi(\mf x)$ depends explicitly on the choice of foliation by Cauchy surfaces $\mathcal{E}_s$ and a time translation direction $s$ that connects the different sheaves, so that it can be written as
    \begin{equation}
        \pi(\mf x) = \frac{\delta S}{\delta (\partial_s \phi(\mf x))}.
    \end{equation}
    Having the momentum associated to this given foliation of spacetime, it is then possible to upgrade $\phi(\mf x)$ and $\pi(\mf x)$ to operators and impose the `equal time' canonical commutation relations
    \begin{align}
    \nonumber    \big[\hat{\phi}(\mf x),\hat{\pi}(\mf x')\big] &= \delta_{\mathcal{E}_s}(\mf x,\mf x')\openone,\\
        \big[\hat{\phi}(\mf x),\hat{\phi}(\mf x')\big] &= 0,\label{2}\\
     \nonumber   \big[{\hat{\pi}(\mf x)},{\hat{\pi}(\mf x')}\big] &= 0,
    \end{align}
    where $\delta_{\mathcal{E}_s}(\mf x,\mf x')$ is the Dirac delta distribution associated to each of the surfaces $\mathcal{E}_s$.
    
    We can then build the usual Fock representation for field states by 
    promoting the coefficients $a_{\bm k}$ and $a_{\bm k}^*$ from Eq. \eqref{modeExp} to operators. This gives rise to the creation and annihilation operators associated to the mode expansion in terms of the basis of solutions $\{u_{\bm k}(\mf x)\}$. That is, the quantum field $\hat{\phi(\mf x)}$ can be written in any point of spacetime as
    \begin{equation}\label{theQField}
        \hat{\phi}(\mf x) = \int \dd^{n} \!\bm{k}\left(\hat{a}^\dagger_{\bm{k}} u_{\bm{k}}^*(\mathsf x)+\hat{a}_{\bm{k}}^{\phantom{\dagger}} u_{\bm{k}}( \mathsf x) \right).
    \end{equation}
    The canonical commutation relations \eqref{2} force the standard bosonic commutation relations for the creation and annihilation operators
    \begin{align}
       \big[\hat{a}_{\bm{k}}^{\phantom{\dagger}},\hat{a}^\dagger_{\bm{k'}}\big] &= \delta^{(n)}(\bm k - \bm k') \openone\nonumber,\\
        \big[\hat{a}_{\bm{k}}^{\phantom{\dagger}},\hat{a}^{\phantom{\dagger}}_{\bm{k'}}\big] &= 0 ,\\
        \big[\hat{a}_{\bm{k}}^{{\dagger}},\hat{a}^\dagger_{\bm{k'}}\big] &= 0.\nonumber
    \end{align}
    With this, a vacuum state $\ket{0}$ associated to this quantization is defined as the state annihilated by all the annihilation operators $\hat{a}_{\bm k}$. The Hilbert space associated to the field is built by successive applications of the creation operators $\hat{a}^\dagger_{\bm k}$ on the vacuum state.
    
    Notice that the choice of orthonormal set $\{u_{\bm k}(\mf x)\}$ is not unique and indeed there are an infinite number of ways of representing the field in terms of a sum of modes. If two different representations are unitarily equivalent then the annihilation operators $\hat{a}_{\bm k}$ and $\hat{a}^\dagger_{\bm k}$ associated to the two different set of modes annihilate the same vacuum. However, even in the simplest scenarios there are non-unitarily equivalent ways of quantizing the field. A typical example is the Rindler quantization. The vacuum associated to a quantization in terms of Minkowski modes corresponds to thermal states in the right and left wedges of a Rindler quantization~\cite{Takagi,birrelldavies}. 
    
    The choice of modes determines the explicit spacetime dependence of the field and therefore fixes its free dynamics. In this manuscript we will assume that this choice has been made at the level of field quantization and the free quantum field is already given as in Eq. \eqref{theQField} for every point of spacetime. It is important to remark that this is a common assumption when using the UDW model in curved backgrounds and it is the assumption that yields covariant predictions for pointlike detectors as we will see.
    
    
    \color{black}
    
   Same as in (among others) \cite{us}, we assume our detector to be localized as a smeared (Fermi-Walker rigid) two-level  first quantized system. We notate  $\mathsf{z}(\tau)$ the trajectory of the detector's centre of mass, parametrized by proper time $\tau$. We denote $\ket{g}$ and $\ket{e}$ the ground and excited state of the detector according to the detector's free Hamiltonian $\hat H_{\text{d}}^\tau$ (which generates translations with respect to $\tau$)
    \begin{equation}\label{hD}
        \hat H_{\textrm{d}}^\tau = \Omega \hat \sigma^{+}\hat \sigma^{-} = \frac{\Omega}{2} \left(\hat{\sigma}_z+\openone\right),
    \end{equation}
    where $\Omega$ is the proper energy gap of the detector and \mbox{$\hat{\sigma}^+=\ket{e}\!\bra{g}=(\hat\sigma^{-})^\dagger$}. 
    
    \color{black} In a given coordinate system  the interaction Hamiltonian density $\hat{\mathfrak{h}}_I(\mf x)$ can be written in terms of a scalar Hamiltonian weight $\hat{h}_I(\mf x)$ according to \mbox{$\hat{\mathfrak{h}}_I(\mf x) = \sqrt{-g}\, \hat{h}_I(\mf x)$}, where $g$ is the determinant of the metric in the corresponding coordinates. This will simplify the analysis because unlike the Hamiltonian density, the Hamiltonian weight is a scalar, and therefore can be used without the need to explicitly mention coordinate systems. The Hamiltonian weight associated to the interaction of a two-level system with a scalar quantum field in the UDW model takes the following shape,
    \color{black}
    
    \begin{equation}
        \hat{h}_I(\barsf{x}) = \lambda  
        \chi(\tau)f(\bar{\bm x})\hat{\mu}(\tau)\hat{\phi}(\bar{\mathsf{x}}).
    \end{equation}
    where---following the prescription from \cite{us}---we pick Fermi normal coordinates $(\tau,\bar{\bm{x}})$, associated to the centre of mass of the detector, and the monopole moment operator takes the form
    \begin{equation}
        \hat \mu(\tau) = e^{\ii\Omega \tau}\hat \sigma^+ +e^{-\ii\Omega\tau}\hat \sigma^-.
        \label{eq:monopole}
    \end{equation}    
    \color{black} $\chi(\tau)$ and $f(\bar{\bm{x}})$ are the switching and smearing functions, respectively. Notice that, by construction, in the proper frame of the detector we can factor a switching function and a spatial smearing function in the interaction Hamiltonian. This is associated to the assumption that the detector  is Fermi-Walker rigid, that is, it keeps its shape in its own reference frame\color{black}. In a general coordinate system there is no factorization of a switching and a smearing function and the Hamiltonian weight will be characterized instead by a spacetime smearing $\Lambda(\mathsf{x})$, that is, 
    \begin{equation}\label{hDensityCov}
        \hat{h}_I(\mf x) = \lambda \Lambda(\mf x) \hat{\mu}(\tau(\mf x))\hat{\phi}(\mf x).
    \end{equation}
    
    As stated in \cite{us}, the integral of the above quantity in spacetime is fully covariant and coordinate independent. The Hamiltonian that generates time evolution with respect to the proper time $\tau$ of the detector is then defined as the integral over the constant $\tau$ surfaces $\Sigma_\tau$, according to
    \begin{equation}\label{HamiltonianTau}
        \hat{H}_{I}^{\tau}(\tau) = \lambda \int_{\Sigma_{\tau}}\!\!\!\dd^n \bar{\bm x} \sqrt{-\bar{g}}\chi(\tau)f(\bar{\bm x})\hat{\mu}(\tau)\hat{\phi}(\bar{\mathsf{x}}),
    \end{equation}
    while the Hamiltonian generating translations with respect to an arbitrary time coordinate $t$ can be written as
    \begin{equation}\label{HamiltonianT}
        \hat{H}_{I}^{t}(t) = \lambda \int_{\mathcal{E}_{t}}\!\!\!\dd^n \bm{x} \sqrt{-{g}}\:\Lambda(\mathsf{x})\hat{\mu}(t)\hat{\phi}({\mathsf{x}}),
    \end{equation}
    where $\mathcal{E}_t$ denotes the constant $t$ spacelike surfaces in the coordinates $\mf x = (t,\bm x)$.
    
    The time evolution operator is then defined as the time-ordered exponential
    \begin{align}\label{ufirst}
        \hat{\mathcal{U}} &= \mathcal{T}_\tau\exp\left(-\ii\!\int_{\mathcal{M}}\!\!\dd\mathcal{V} \hat{h}_I(\mf x)\right)= \mathcal{T}_\tau\exp\left(-\ii\!\int_\mathbb{R}\!\!\dd\tau \hat{H}_I^\tau(\tau)\right),
    \end{align}
    where \textcolor{black}{we have made it explicit that} the time ordering operator $\mathcal{T}_\tau$ represents time ordering with respect to the proper time of the detector's centre of mass $\tau$.

    
    

\color{black}
It is important to notice that the UDW model has historically been prescribed at the Hamiltonian level, and not from a `first-principle' action. It is indeed an effective model built to bypass the need for a full relativistic description of the detector's internal dynamics. It is nevertheless possible to obtain the interaction Hamiltonians in~\eqref{HamiltonianTau} and \eqref{HamiltonianT} from the following interaction \emph{action}:
\begin{equation}\label{interactionaction}
    S_I = \int \dd^D\bar{\mathsf{x}} \sqrt{-\bar{g}} \mathcal{L}_I(\bar{\mathsf{x}}),
\end{equation}
where $\mathcal{L}_I$ is a scalar interaction Lagrangian weight. From this action, an interaction energy-momentum tensor can be assigned:
\begin{equation}
    T^{ab}_I = -\dfrac{2}{\sqrt{-g}}\dfrac{\delta S_I}{\delta g_{ab}}.
\end{equation}
Now assume that the interaction Lagrangian does not explicitly depend on the metric. This is true for common potential and interaction terms in scalar field theories and certainly true for the common UDW detector models as introduced in previous literature~\cite{Unruh1976,DeWitt,Unruh-Wald}. Then, the only dependence on the metric in~\eqref{interactionaction} comes from the volume element, so that we obtain
\begin{equation}
    T_I^{ab} = -\mathcal{L}_I(\bar{\mathsf{x}}) g^{ab}.
\end{equation}
The interaction Hamiltonian~\eqref{interactionhamiltonian} can then be obtained as
\begin{equation}\label{interactionhamiltonian2}
    H^\tau_I = \int_{\Sigma_\tau} \dd \Sigma\,\,  n_a T_I^{ab} \xi_b,
\end{equation}
where $\xi^a = (\partial_\tau)^a$, $\dd\Sigma$ is the induced volume element on the spacelike surfaces $\Sigma_\tau$ with unit normal $n^a$, and we complete the identification from~\eqref{interactionhamiltonian2} to~\eqref{interactionhamiltonian} by recognizing $h_I(\bar{\mathsf{x}}) = -\mathcal{L}_I(\bar{\mathsf{x}})$.

We will take as an \emph{assumption} of the setup that, once the free quantization of the field has been performed, its spacetime dependence in the interaction picture is fully determined. In particular, this implies that any local interaction term in an interaction action between the fields in our setup will correspond to an interaction Hamiltonian weight $\hat{h}_I(\mathsf{x})$ that is a \emph{foliation-independent} scalar function of the coordinates. This is a nontrivial assumption, insofar as it is hard to justify it in general from first-principle arguments based on a careful description of the free and interacting Hamiltonians arising from splitting the full energy-momentum tensor of the theory as a ``free'' and an ``interacting'' part. Nevertheless, this has certainly been taken for granted in the standard approaches based on particle detector models with detectors in trajectories that may not correspond to the trajectories of fiducial observers according to which canonical quantization (and in particular, the definition of the vacuum) has been performed. Notice again that this choice yields covariant predictions for pointlike detectors. Since the model has been remarkably successful in QFT in many scenarios where the non-triviality manifests (for example predicting the Unruh effect~\cite{Takagi,Unruh-Wald,Unruh1976}) and our our objective is to highlight the limitations that are intrinsic to the usual strategy employed in smeared particle detector models. We find this a fair assumption upon which to base our following remarks.
    \color{black}
\section{The Time Ordering Operation}\label{timeOrderSection}

The notion of time ordering is fundamental in our understanding of time evolution in quantum  theory.  When a given coordinate system is chosen, $\mf x = (t,\bm x)$, the time ordering of events associated to this coordinate system is understood as an ordering with respect to the coordinate time $t$. For timelike or null separated events, time ordering is independent of the coordinate system picked. However, for spacelike events this is not the case. In this section, we will study under which conditions the time-ordered exponential of a Hamiltonian density is independent of the time parameter used to order it. \color{black}We will do so for a scalar quantum field theory in a globally hyperbolic spacetime $\mathcal{M}$ of dimension $D = n+1$  with metric $g$, in the context of the UDW model discussed in Section \ref{UDWmodel}.

Consider the UDW model, as presented in Section \ref{UDWmodel}. In the Fermi normal coordinates associated to the detector's center of mass worldline, we have seen that the Hamiltonian and unitary time evolution operator are respectively given by 
\color{black}
\begin{equation}
    \hat{H}_I^\tau(\tau) = \int \dd^n \bar{\bm{x}} \sqrt{-\bar{g}}\, \hat{h}_I(\bar{\mf x}),\label{interactionhamiltonian}
\end{equation}
\vspace{-15pt}
\begin{equation}      
    \hat{\mathcal{U}}_\tau = \mathcal{T}_\tau\exp\left(-\ii\int_\mathbb{R}\! \dd\tau\: \hat{H}_I^\tau(\tau) \right).
\end{equation}
\color{black}
\textcolor{black}{This time evolution operator should then be thought to evolve initial data, encoded in general as a state operator $\hat{\rho}_0$ and prescribed at an initial Cauchy surface $\Sigma_{\tau_0}$, to a final future Cauchy surface $\Sigma_{\tau_1}$. Throughout our discussion, we will in general assume $(\tau_0, \tau_1) \rightarrow (-\infty, +\infty)$. Notice that any finite nature of the interaction would be implemented through the possibly finite spacetime support in the Hamiltonian. }

\textcolor{black}{Under the assumptions outlined in Section \ref{UDWmodel}, one can alternatively compute the time evolution prescribed by a different coordinate time $t$ by assigning an interaction Hamiltonian given by}
 \begin{equation}\label{hCovGeneral}
        \hat{H}^{t}_I(t) = \int_{\mathcal{E}_{t}} \!\! \dd^n \bm{x} \sqrt{-g}\:\hat{h}_I(\mathsf{x}),
    \end{equation}
where $\bm{x}$ are spacelike coordinates on $\mathcal{E}_{t}$, which are the surfaces of simultaneity defined by constant $t$. \textcolor{black}{When comparing the time evolution generated by~\eqref{interactionhamiltonian} and~\eqref{hCovGeneral}, one should keep in mind that we are implicitly assuming that the past and future Cauchy surfaces corresponding to $(t_0, t_1)$ coincide with the ones associated to $(\tau_0, \tau_1)$---otherwise, the comparison would be meaningless, since it would involve comparing observables located in different spatial slices. Again, this does not mean that we cannot model finite-time interactions since the finiteness will be encoded in the spacetime support of $ \hat{H}^{t}_I(t)$}

Having a covariantly defined Hamiltonian as in \eqref{hCovGeneral} is, however, not enough to guarantee that the time evolution operator itself will be independent of the time parameter chosen to prescribe it. This will only be true if the time ordering operation with respect to $\tau$ is actually truly independent of the time coordinate chosen. If this were not the case, it is easy to see that issues with time-ordering will appear in every order $\mathcal{O}(\lambda^n)$ with $n\geq 2$ of the Dyson expansion of $\hat{\mathcal{U}}_\tau$. Namely, if we write the Dyson expansion as
\begin{equation}
    \hat{\mathcal{U}}_\tau=\openone+\hat{\mathcal{U}}^{(1)}_\tau+ \hat{\mathcal{U}}^{(2)}_\tau +\mathcal{O}(\lambda^3),
\end{equation}
then the time ordering prescription $\mathcal{T}_\tau$ associated to the the detector's proper time yields for the second order term
\begin{align}
   \nonumber \hat{\mathcal{U}}^{(2)}_\tau &\coloneqq (-\ii)^2 \int_{-\infty}^{+\infty}\!\!\!\dd \tau \int_{-\infty}^{\tau} \!\!\!\dd\tau'\, \hat{H}^{\tau}_I(\tau)\hat{H}^{\tau'}_I(\tau')\\
    &=(-\ii)^2\int_{\mathcal{M}\times\mathcal{M}} \!\!\!\!\!\!\!\!\!\!\!\!\dd \mathcal{V} \dd \mathcal{V}'\: \hat{h}_I(\bar{\mathsf{x}})\hat{h}_I(\bar{\mathsf{x}}')\theta(\tau-\tau'),
    \label{U2f}
\end{align}
where $\theta(\tau)$ is the Heaviside step function.
    
 If we now try to perform a coordinate transformation to another coordinate system  \mbox{$\mathsf{x} = (t, \bm{x})$}, we get 
    \begin{align}\label{U2tau}
      \hat{\mathcal{U}}^{(2)}_\tau&=  (-\ii)^2\int_{\mathcal{M}\times \mathcal{M}} \!\!\!\!\!\!\!\!\!\!\!\!\dd \mathcal{V} \dd \mathcal{V}'\: \hat{h}_I({\mathsf{x}})\hat{h}_I({\mathsf{x}'})\theta\big(\tau(\mathsf{x})-\tau'(\mathsf{x}')\big)\\
      &\neq (-\ii)^2 \int_{-\infty}^{+\infty}\!\!\!\dd  t  \int_{-\infty}^{ t } \!\!\!\dd t '\, \hat{H}^{ t }_I( t )\hat{H}^{ t '}_I( t ') = \hat{\mathcal{U}}^{(2)}_t. \nonumber
    \end{align}
    Because there can be spacelike separated events in the integral in \eqref{U2tau}, we do not get the nested integration that one would expect from carrying out time ordering $\mathcal{T}_t$ with respect to the time coordinate $t$ instead of $\tau$. 
    
    Note, however, that we can split the integration region $\mathcal{M}\times \mathcal{M}$ into four subregions:
    \begin{align}
        &\!\!\!\!T :=\{(\bar{\mathsf{x}},\bar{\mathsf{x}}')\in \mathcal{M}\!\times\! \mathcal{M}: \barsf{x} \timelike 
        \barsf{x}' \},\\
        &\!\!\!\!N :=\{(\bar{\mathsf{x}},\bar{\mathsf{x}}')\in \mathcal{M}\!\times\! \mathcal{M}: \barsf{x} \lightlike \barsf{x}' \},\\
        &\!\!\!\!S_> := \{(\bar{\mathsf{x}},\bar{\mathsf{x}}')\in \mathcal{M}\!\times\! \mathcal{M}: \barsf{x} \spacelike \barsf{x}' \textrm{ and }  \tau\!>\!\tau' \Rightarrow t\!>\! t' \},\\
        &\!\!\!\!S_\le :=  \{(\bar{\mathsf{x}},\bar{\mathsf{x}}')\in \mathcal{M}\!\times\! \mathcal{M}: \barsf{x} \spacelike \barsf{x}' \textrm{ and }  \tau\!>\!\tau' \Rightarrow t\!\le \!t' \},
    \end{align}
    where $\mathsf{x}\timelike \mathsf{x'}$ 
    corresponds to timelike, $\mf x \lightlike \mf x'$ corresponds to null 
    and $\mathsf{x}\spacelike \mathsf{x'}$ corresponds to spacelike separation between $\mathsf{x}$ and $\mathsf{x}'$. 
    
    With this splitting we can write $\hat{\mathcal{U}}^{(2)}$ in \eqref{U2f} as a sum of integrals over the different regions 
    \begin{align}
        \hat{\mathcal{U}}^{(2)}_\tau \!
        &=(-\ii)^2\int_{T\,\cup\, N\,\cup\, S_>}\!\!\!\!\!\!\!\!\!\!\!\!\!\!\!\!\!\!\! \dd \mathcal{V} \dd \mathcal{V}'\,
        \hat{h}_I(\bar{\mathsf{x}})\hat{h}_I(\bar{\mathsf{x}}')\theta(\tau\!-\!\tau')\nonumber\\
        &+(-\ii)^2\int_{S_\leq}\!\!\dd \mathcal{V} \dd \mathcal{V}'\: \hat{h}_I(\bar{\mathsf{x}})\hat{h}_I(\bar{\mathsf{x}}')\theta(\tau\!-\!\tau').
        \label{U2s}
    \end{align}
For timelike and null separation, the time ordering between two events is the same for every observer, which means that for points on regions $T$ and $N$,  $\tau(\mathsf{x})-\tau'(\mathsf{x}') > 0 \iff t - t' > 0$ as per the discussion in Section \ref{intervals}. This allows us to equate $\theta\big(\tau(\mathsf{x})-\tau(\mathsf{x}')\big) = \theta(t - t')$ in these regions. 
The same reasoning is true for the points in the $S_>$ region by construction, since we defined $S_>$ to be the region composed of spacetime events that preserved the previous time ordering. 

The only region where the coordinate transformation may cause problems is $S_\leq$, since it changes the time ordering between the two events. In this region, we can write $\theta\big(\tau(\mathsf{x})-\tau(\mathsf{x}')\big) = \theta(t' - t)$, which allows us to rewrite the integral as
    \begin{equation}
    \begin{gathered}
        \int_{S_\leq} \!\!\!\dd \mathcal{V} \dd \mathcal{V}' \hat{h}_I({\mathsf{x}})\hat{h}_I({\mathsf{x}'})\theta\big(\tau(\mathsf{x})-\tau(\mathsf{x}')\big)\\
        =\int_{S_\leq} \!\!\!\dd \mathcal{V} \dd \mathcal{V}' \hat{h}_I({\mathsf{x}})\hat{h}_I({\mathsf{x}'})\theta\big(t' - t\big);
    \end{gathered}
    \end{equation}
    Then, writing $\hat{h}_I({\mathsf{x}})\hat{h}_I({\mathsf{x}'}) = \hat{h}_I({\mathsf{x}'})\hat{h}_I({\mathsf{x}}) + [\hat{h}_I({\mathsf{x}}), \hat{h}_I({\mathsf{x}'})]$, we get
    \begin{equation}
        \begin{gathered}
        \int_{S_\leq} \!\!\!\dd \mathcal{V} \dd \mathcal{V}' \hat{h}_I({\mathsf{x}})\hat{h}_I({\mathsf{x}'})\theta\big(t' - t\big)\\
        = \int_{S_\leq} \!\!\!\dd \mathcal{V} \dd \mathcal{V}'  \hat{h}_I({\mathsf{x}'})\hat{h}_I({\mathsf{x}})\theta\big(t' - t\big)\\
        +\int_{S_\leq} \!\!\!\dd \mathcal{V} \dd \mathcal{V}' [\hat{h}_I({\mathsf{x}}), \hat{h}_I({\mathsf{x}'})]\theta\big(t' - t\big).
        \end{gathered}
    \end{equation}
    Renaming the integration variables $\mathsf{x}$ and $\mathsf{x}'$ in the first integral of the right hand side above we recover the same integrand as in Eq. \eqref{U2tau}. Adding the integrals over the regions 
    ${T},N, S_>$ and $S_\leq$, we finally get
    \begin{align}
       \!\!\!\!  \hat{\mathcal{U}}^{(2)}_\tau &=(-\ii)^2\int_{\mathcal{M}\times\mathcal{M}} \!\!\!\!\!\!\!\!\!\!\!\!\dd \mathcal{V} \dd \mathcal{V}' \hat{h}_I(\bar{\mathsf{x}})\hat{h}_I(\bar{\mathsf{x}}')\theta(\tau-\tau')\nonumber\\
      \!\!\!\!\!   &=(-\ii)^2\int_{\mathcal{M}\times\mathcal{M}} \!\!\!\!\!\!\!\!\!\!\!\!\dd \mathcal{V} \dd \mathcal{V}' \hat{h}_I(\mathsf{x})\hat{h}_I(\mathsf{x}')\theta(t - t')\nonumber\\
      \!\!\!\!\!   &+(-\ii)^2\int_{S_\leq} \!\!\!\!\!\dd \mathcal{V} \dd \mathcal{V}' [\hat{h}_I({\mathsf{x}}), \hat{h}_I({\mathsf{x}'})]\theta\big(t' - t\big)\nonumber\\
      \!\!\!\!\!   &= \hat{\mathcal{U}}^{(2)}_t+(-\ii)^2\int_{S_\leq} \!\!\!\!\!\dd \mathcal{V} \dd \mathcal{V}' [\hat{h}_I({\mathsf{x}}), \hat{h}_I({\mathsf{x}'})]\theta\big(t' - t\big),\label{U2correction}
   \end{align}
    where we recall $\mathcal{T}_t$ represents time ordering with respect to $t$ and $\hat{\mathcal{U}}_t$ the associated time evolution operator. The second summand in \eqref{U2correction}   ultimately threatens the covariance of the time ordering prescription. This term is proportional to the commutator of the Hamiltonian densities at spacelike-separated points. 
    
    To generalize the result above to higher orders, notice that the $N$th term in the Dyson series can be written as
    \begin{align}
        \nonumber\hat{\mathcal{U}}^{(N)}_\tau &= \dfrac{(-\ii)^N}{N!}\!\!\int_{\mathcal{M}^N}\!\!\!\!\!\!\dd \mathcal{\mathcal{\mathcal{V}}}_1\dots \dd \mathcal{V}_N\:\mathcal{T}_\tau \hat{h}_I(\bar{\mathsf{x}}_1)\dots\hat{h}_I(\bar{\mathsf{x}}_N)\\
        &= \dfrac{(-\ii)^N}{N!}\!\!\int_{\mathcal{M}^N}\!\!\!\!\!\!\dd \mathcal{\mathcal{\mathcal{V}}}_1\dots \dd \mathcal{V}_N\:\mathcal{T}_\tau \hat{h}_I({\mathsf{x}}_1)\dots\hat{h}_I({\mathsf{x}}_N)\label{U2}\\
        &\neq \dfrac{(-\ii)^N}{N!}\!\!\int_{\mathcal{M}^N}\!\!\!\!\!\!\dd \mathcal{\mathcal{\mathcal{V}}}_1\dots \dd \mathcal{V}_N\:\mathcal{T}_t \hat{h}_I({\mathsf{x}}_1)\dots\hat{h}_I({\mathsf{x}}_N).\nonumber
    \end{align}    
    where $\mathcal{T}_\tau$ applied to the Hamiltonian densities time-orders the product according to the detector's centre of mass proper time $\tau$. As the second and third line of \eqref{U2} shows, we could switch from the $\bar{\mf x} = (\tau,\bar{\bm x})$ coordinates to arbitrary coordinates $\mf x = (t,\bm x)$ without picking any extra terms, but we need to keep the time ordering with respect to $\tau$. Expressing the time-ordering with respect to $\tau$ in terms of time ordering in the coordinates $(t,\bm x)$ is, in general, a nontrivial task but it is in general different from time ordering with respect to $t$.
    
We would like to highlight that the non-coincidence of time-ordering with respect to different coordinate systems can be bypassed in many common scenarios. If the Hamiltonian weight is microcausal, that is, it satisfies 
\begin{equation}\label{microcausality}
\mathsf{x}\spacelike\mathsf{x}' \Rightarrow   [\hat{h}(\mf x),\hat{h}(\mf x')] = 0,
\end{equation}
then the ambiguity in time ordering of spacelike-separated events will have no impact in the calculation of the time evolution operator. In other words, when \eqref{microcausality} is satisfied, the time evolution operator is the same with respect to any time parameter. There are many relevant interactions where the Hamiltonian density is microcausal. The postulate of microcausality in QFT implies that field operators evaluated at spacelike-separated points commute. If the interaction Hamiltonian weight  $\hat{h}_I(\mf x)$ is local in the quantum fields (that is, only couples the detector to field degrees of freedom evaluated at a single point in each spatial slice) the Hamiltonian weight is microcausal. This is why in (most of) high-energy physics, where all fields are microcausal and the interactions are local, there is no need to specify a privileged time ordering and the time evolution is always covariant. This is also why a detection scheme based on the Fewster-Verch QFT measurement framework~\cite{fewster1,fewster2,fewster3} would not have any problems with covariance. However, smeared particle detectors such as the smeared UDW model involve non-local couplings to quantum fields, hence will suffer from time-ordering ambiguities as we will see.

\section{Breaking of Covariance by a single smeared detector}\label{CovOneDetector}
    


    In section \ref{timeOrderSection} we have seen that local quantum field theories that satisfy microcausality (observables commute at different spacelike separated points) would produce time evolution operators that do not depend on the time parameter chosen for time ordering. When we take $\hat{h}_I(\mf x)$ to be the Hamiltonian weight  associated to a single pointlike detector undergoing an arbitrary timelike trajectory in a fixed background, the interaction is local. That is, the detector's degree of freedom only couples to a single point in each space slice. This translates into the fact that the support of the Hamiltonian weight  $\hat{h}_I(\mf x)$ consists of a single point in each spatial slice, which in turn implies that $\hat{h}_I(\mf x)$ satisfies a microcausality condition: it commutes with itself at spacelike separated points. In summary: predictions of the time evolution of  pointlike UDW detectors coupled through Hamiltonian weights of the form \eqref{hDensityCov} are fully covariant.
    

    However, in the smeared UDW detector model \color{black} employed in the literature, the Hamiltonian weight (and therefore the Hamiltonian density) can be shown to violate microcausality. Namely, the commutator of the UDW interaction Hamiltonian densities for a single smeared detector evaluated at spacelike-separated points is not identically zero due to the smearing. To the authors' knowledge, this has not been taken into account in previous literature. This represents an important limitation of the theory that must be accounted for even for the simplest cases of inertial detectors in flat spacetimes (as we will show in an example later). This violation of general covariance can then be used to stipulate a limit of validity and better clarify under which conditions the predictions of particle detector models can be trusted. This violation (that would yield coordinate dependence of the model's predictions) is deeply linked  to the fact that the spatially smeared UDW Hamiltonian itself encodes an interaction of a single degree of freedom of the detector with a field observable in a region with finite spatial extension, and is therefore inherently nonlocal. 
    The goal of this subsection is therefore to quantify the degree to which this nonlocality of the interaction hinders the covariant nature of predictions prescribed in different coordinate systems. In other words, evaluate how good an approximation we are taking when we consider a smeared UDW detector to model the underlying covariant theory describing the interaction of field and detectors.\color{black}
    
    To quantify the break of covariance introduced by the smearing we make use of the results of Section \ref{timeOrderSection}, by taking the coordinates $\bar{\mf x} = (\tau,\bar{\bm x})$ to be the Fermi normal coordinates associated to the detector's center of mass and we take $\mf x = (t,\bm x)$ to be a different arbitrary frame. We recall that time ordering is unambiguous for the timelike and null regions $T$ and $N$. Furthermore the only region where time ordering can cause covariance problems is $S_\leq$ since, by definition, it contains all the events for which time-ordering is not the same in both frames. Considering that the quantum field theory satisfies microcausality ($[\hat{\phi}(\mf x),\hat{\phi}(\mf x')] = 0$ for $\mathsf{x}\spacelike\mathsf{x}'$), we can write the commutator of the Hamiltonian weights in \eqref{U2correction} in terms of the commutator of the monopole operator at different times in  $S_{\leq}$ as
    \begin{equation}\label{commhI}
        \!\big[\hat{h}_I(\mathsf{x}),\hat{h}_I(\mathsf{x}')\big]\!=\!  \lambda^2 \Lambda(\mathsf{x})\Lambda(\mathsf{x}') \big[\hat{\mu}(\tau(\mathsf{x})),\hat{\mu}(\tau(\mathsf{x}'))\big]  \hat\phi(\mathsf{x})\hat \phi(\mathsf{x'}),
    \end{equation}
    where the $\Lambda(\mathsf{x})$ is the spacetime smearing function. From \eqref{eq:monopole} we can explicitly evaluate the monopole moment commutator for a qubit UDW detector as
    \begin{equation}
        \big[\hat{\mu}(\tau),\hat{\mu}(\tau')\big] = 2\ii \sin(\Omega(\tau-\tau'))\hat{\sigma}_z.\label{commMu}
    \end{equation}
    \color{black} Notice that this commutator vanishes only for specific times, namely, when $\tau = \tau'+\pi n /\Omega$ for integer values of $n$, hence the smeared UDW detector interaction Hamiltonian density breaks microcausality. It is important to remark that this issue is present even in the simplest scenarios already studied in the literature, such as inertial motion of particle detectors in flat spacetimes. 
    
    As mentioned in Section \ref{timeOrderSection}, when an interaction Hamiltonian does not satisfy microcausality, the time ordering associated to different notions of time translations might impact the result for the time evolution operator. This implies that the uses of smeared particle detectors not relying on a quantum field theoretical description of the detector (i.e., particle detector models except for the Fewster-Verch approach~\cite{fewster1,fewster2,fewster3})  implicitly assume a notion of time translation with respect to which the calculations are performed, and, in principle, different choices of such notions might have yielded different predictions. However, a mathematical model that represents reality cannot yield different results depending on the coordinate system used to perform computations. It is thus important to quantify the difference that the different choices of coordinates introduce in the time evolution operator. If these differences were to be relevant, they may cast doubt on the accuracy of the predictions made by particle detector models.
    
    To study the possible coordinate dependence of the predictions of particle detector models let us start by comparing the time ordering operation associated to the proper time of the detector's frame $\tau$ with the time ordering with respect to a different parameter $t$, associated to a foliation $\mathcal{E}_t$. We make use of the result of Eq. \eqref{commhI}, where the commutator of $\hat{h}_I(\mf x)$ with itself at different events was explicitly evaluated. \color{black} Concretely, consider two time-ordered exponentials that define two different time evolution operators $\hat{\mathcal{U}}_\tau$ and $\hat{\mathcal{U}}_t$. On the one hand, $\hat{\mathcal{U}}_\tau$ is associated to the Hamiltonian generating  time evolution with respect to the proper time of the detector $\hat{H}_I^\tau(\tau)$, that is
    \begin{align}\label{u30}
        \hat{\mathcal{U}}_\tau = \mathcal{T}_\tau\exp\left(-\ii\! \int\!\dd \tau\, \hat{H}_I^\tau(\tau)\right).
    \end{align}
     On the other hand, the time evolution operator $\hat{\mathcal{U}}_t$ is associated to the time-order of the Hamiltonian $\hat{H}_I^t(t)$ generating translations with respect to another time parameter $t$:
    \begin{align}
        \hat{\mathcal{U}}_t = \mathcal{T}_t\exp\left(-\ii\! \int\!\dd t\, \hat{H}_I^t(t)\right).
    \end{align}
    In a covariant formalism we should have $\hat{\mathcal{U}}_\tau=\hat{\mathcal{U}}_t$, so that the predictions do not depend on the choice of coordinates.
    \color{black}
    While this is not going to be the case for non-pointlike detectors, it is possible to precisely quantify the difference between the two time-ordering prescriptions in a general smearing scenario. \color{black} As discussed in Section \ref{timeOrderSection}, for the first order Dyson expansion term in the time evolution, we verify that $\hat{\mathcal{U}}^{(1)}_\tau = \hat{\mathcal{U}}^{(1)}_t$ and therefore the first deviation from using the two coordinate systems appears in the second order of the Dyson expansion. From \eqref{U2correction} we get
    \begin{equation}\label{deviationOne}
        \hat{\mathcal{U}}_t^{(2)} - \hat{\mathcal{U}}_\tau^{(2)} = - \int_{S_\leq} \!\!\!\dd \mathcal{V} \dd \mathcal{V}' \comm{\hat{h}_I({\mathsf{x}})}{\hat{h}_I({\mathsf{x}'})}\theta\big(t' - t\big).
    \end{equation}
    If we expand the integral above using the expression of the Hamiltonian weight  $\hat{h}_I(\mathsf{x})$ in terms of the field and monopole operators and equation \eqref{commMu}, we obtain
   \begin{align}
        \nonumber \hat{\mathcal{U}}_t^{(2)} - \hat{\mathcal{U}}_\tau^{(2)}=- 2 \ii \lambda^2 \hat{\sigma}_z\!\!&\int_{S_\leq} \!\!\!\!\!\dd \mathcal{V} \dd \mathcal{V}'\Lambda(\mathsf{x})\Lambda(\mathsf{x}')\hat\phi(\mathsf{x})\hat \phi(\mathsf{x'})\\
        &\times\sin\big[\Omega(\tau-\tau')\big]\theta\big(t' - t\big).
    \end{align}
    
    We then define an  operator $\hat{E}$ that acts only on the Hilbert space of the field as
    \begin{align}\label{defE}
        \hat{E}\!\coloneqq\! \!- 2 \ii\!\!\int_{S_\leq} \!\!\!\!\!\dd \mathcal{\mathcal{V}} \dd \mathcal{V}'\!\Lambda(\mathsf{x})\Lambda(\mathsf{x}')\hat\phi(\mathsf{x})\hat \phi(\mathsf{x'}) \sin\!\big[\Omega(\tau\!-\!\tau')\big]\theta\big(t'\!\! - \!t\big),
    \end{align}
    so that we can write the difference between $\hat{\mathcal{U}}_t^{(2)}$ and $\hat{\mathcal{U}}_\tau^{(2)}$  as
    \begin{equation}\label{jonas}
        \hat{\mathcal{U}}_t^{(2)} - \hat{\mathcal{U}}^{(2)}_\tau = \lambda^2 \hat{\sigma}_z \hat{E}.
    \end{equation}
    Taking the adjoint of Equation \eqref{defE} and using the fact that the field operators commute when evaluated at points in $S_\leq$, one sees that $\hat{E}^\dagger = -\hat{E}$.
   
    We can evaluate the exact magnitude of the violation of covariance by choosing a particular initial state for detector and field. In particular, in the reasonable scenario that field and detector are initially uncorrelated,  the initial joint state is 
    \begin{equation}
        \hat{\rho}_0 = \hat{\rho}_{\text{d},0}\otimes \hat{\rho}_\phi.
    \end{equation}
    After the  interaction, the state of the field-detector system will be given by
    \begin{equation}
        \hat{\rho}^\tau = \hat{\mathcal{U}}_\tau\hat{\rho}_0 \hat{\mathcal{U}}_\tau^\dagger.
    \end{equation}
    The time-evolved state of the detector is obtained after tracing over the field degrees of freedom:  \mbox{$\hat{\rho}_\textrm{d} = \Tr_\phi \hat{\rho}$}.

    If one decides to prescribe the interaction using any other coordinate system, general covariance would demand that the time evolution implemented by $\hat{\mathcal{U}}_t$ should coincide with that of $\hat{\mathcal{U}}_\tau$. For $\hat{\mathcal{U}}_t$, the density operator used to describe the system after the interaction will be given by
    \begin{equation}
        \hat{\rho}^t = \hat{\mathcal{U}}_t \hat{\rho}_0 \hat{\mathcal{U}}_t^\dagger.
    \end{equation}
   Since the spacetime region of interaction is given by the support of the spacetime profile $\Lambda(\mf x)$ which is coordinate invariant, we can then use Equation \eqref{jonas} to compare $\hat{\rho}^{t}$ with $\hat{\rho}^\tau$. We obtain
    \begin{align}
       \nonumber \hat{\rho}^t &= \hat{\mathcal{U}}_\tau \hat{\rho}_0 \hat{\mathcal{U}}_\tau^\dagger + \lambda^2\left( \hat{\sigma}_z \hat{E} \hat{\rho}_0 \hat{\mathcal{U}}_\tau^\dagger+\hat{\mathcal{U}} _\tau\hat{\rho}_0 \hat{\sigma}_z\hat{E}^\dagger\right)+\mathcal{O}(\lambda^3)\\
        & = \hat{\rho}^\tau +\lambda^2\left(\hat{\sigma}_z\hat{\rho}_{\textrm{d},0}\otimes\hat{E}\hat{\rho}_\phi+\hat{\rho}_{\textrm{d},0}\hat{\sigma}_z\otimes \hat{\rho}_\phi \hat{E}^\dagger\right) + \mathcal{O}(\lambda^3).
    \end{align}
    The covariance breaking introduced in the detector evolved states can be evaluated by partial-tracing the field.
    Using the cyclic property of the trace and that $\hat{E}=-{\hat{E}}^\dagger$  we can  write $\hat \rho^t_{\text{d}}= \Tr_\phi \hat{\rho}^t$ as
    \begin{align}
        \hat{\rho}^t_\textrm{d} &= \hat{\rho}_\textrm{d}^\tau + \lambda^2\left( \hat{\sigma}_z\hat{\rho}_{\textrm{d},0}\Tr \hat{E}\hat{\rho}_\phi+\hat{\rho}_{\textrm{d},0}\hat{\sigma}_z\Tr \hat{\rho}_\phi \hat{E}^\dagger\right)+\mathcal{O}(\lambda^3)\nonumber\\
        &= \hat{\rho}^\tau_\textrm{d} + \lambda^2\left( \hat{\sigma}_z\hat{\rho}_{\textrm{d},0}\Tr \hat{E}\hat{\rho}_\phi-\hat{\rho}_{\textrm{d},0}\hat{\sigma}_z\Tr \hat{\rho}_\phi \hat{E}\right)+\mathcal{O}(\lambda^3)\nonumber\\
        &= \hat{\rho}_\textrm{d}^\tau + \lambda^2\comm{\hat{\sigma}_z}{\hat{\rho}_{\textrm{d},0}}\Tr \hat{\rho}_\phi \hat{E}+\mathcal{O}(\lambda^3),
    \end{align}
    where $\Tr \hat{\rho}_\phi\hat{E}$ can be written in terms of the field state Wightman function as
    \begin{align}\label{citeE}
        \Tr \hat{\rho}_\phi\hat{E}\!=- 2 \ii\!\!\int_{S_\leq} \!\!\!\!\!\dd \mathcal{\mathcal{V}} \dd \mathcal{V}'\!\Lambda(\mathsf{x})&\Lambda(\mathsf{x}')W_{\hat{\rho}_\phi}(\mf x,\mf x')\\& \times\sin\!\big[\Omega(\tau\!-\!\tau')\big]\theta\big(t'\!\! - \!t\big).\nonumber
    \end{align}
    We therefore obtain that the difference between both descriptions is given by
    \begin{equation}\label{podeSIM}
        \hat{\rho}_\textrm{d}^t-\hat{\rho}_\textrm{d}^\tau =  \lambda^2\comm{\hat{\sigma}_z}{\hat{\rho}_{\textrm{d},0}}\Tr \hat{\rho}_\phi \hat{E}+\mathcal{O}(\lambda^3).
    \end{equation}\color{black}
    Equation \eqref{podeSIM} quantifies how much the standard smeared UDW particle detector model changes if one decides to perform the calculations in another reference frame. It is important to remark that the predictions of such models can only be trusted up to the point where the difference between the states $\hat{\rho}_\textrm{d}^t$ and $\hat{\rho}_\textrm{d}^\tau$ is negligible. For example, smeared UDW models provide an accurate approximation for the description of a probe coupling to a quantum field up to second order if the initial state of the detector commutes with its free Hamiltonian, a common choice in many previous works. 
    
    Recall that we discussed in Section~\ref{timeOrderSection} that the break of covariance is linked to the non-local coupling of a single quantum degree of freedom of the detector to multiple spacelike separated points. Indeed,  we can check from Eq. \eqref{podeSIM} that $\hat{E}$ is identically zero for a pointlike detector due to the fact that the pointlike smearing (a delta function) has no points in the region $S_\leq$.  \color{black} Furthermore, if the initial state of the field is a Gaussian state with vanishing one-point function---i.e., \mbox{$\langle \hat\phi(\mathsf{x})\rangle_{\hat{\rho}_\phi}=0$}, which includes not only the vacuum or any thermal state but also any squeezed thermal state---there is no breakdown of covariance even at order $\mathcal{O}(\lambda^3)$. This can be seen by noting that the corrections of order $\lambda^3$ are proportional to integrals of the three point function $\langle{\hat{\phi}(\mathsf{x_1})\hat{\phi}(\mathsf{x_2})\hat{\phi}(\mathsf{x_3})}\rangle_{\hat{\rho}_\phi}$, which is zero for any Gaussian state with vanishing one-point function.


    \color{black}
    It is also worth to point out that in the cases where there is violation of covariance at leading order, this violation is due to the smearing of the detector and is therefore suppressed with the  smearing decay in spacetime as can be seen from Eq. \eqref{defE}. \color{black} This is congruent with  the causality violations found in early literature associated to the smearing of particle detectors \cite{martin-martinez2015}. There, the causality violations were  deemed controllable if they decayed at least as fast as the detector's smearing function tails. \color{black}  Therefore, in the event where the predictions of the model are  taken for proper timescales and lengthscales much larger than the light-crossing time of the detector's smearing, the difference between the two time-ordered evaluations should be negligible when the frames are related by non-extreme accelerations and curvatures, providing regimes for which the model's violation of covariance is negligible. These regimes are precisely the regimes where using particle detectors is meaningful according to other relativistic considerations~\cite{us}, and are well within the regimes where phenomena such as the Unruh effect should become observable. We explicitly illustrate this with an example in Section \ref{ExampleFlatSpacetime}.
    \color{black}
    \section{Covariance of multiple detectors}
    
    After the analysis of the covariance violations in the time-ordering for a single detector, one can wonder what happens when we have scenarios with multiple detectors where, arguably, covariance could be more subtle. These scenarios can combine detectors whose proper times are radically different, and where identifying regimes of timelike or spacelike separation between them is crucial (for example in entanglement harvesting \cite{Valentini1991,Reznik1,reznik2,Retzker2005,RalphOlson1,RalphOlson2,Nick,Cosmo,Salton:2014jaa,Pozas-Kerstjens:2015}). 

    One can work out the deviation between the time-evolution operators defined by different time coordinates in the case of multiple detectors as a straightforward generalization of what was done in Section \ref{CovOneDetector}. Assume we have $N$ detectors, labelled by $j = 1,\dots, N$ whose centres of mass undergo trajectories $\mf z_j(\tau_j)$ parametrized by the proper time of each detector's center of mass, $\tau_{j}$.
    We then prescribe the interaction Hamiltonian densities (or equivalently their weights) in the Fermi normal coordinates associated to each of the detectors' worldlines, $\bar{\mf x}_{j} = (\tau_{j},\bar{\bm x}_{j})$, according to
    \begin{equation}
        \hat{h}_{I,j}(\bar{\mf x}_{j}) = \lambda_j \chi_j(\tau_j)\hat{\mu}_j(\tau_{j}) f_j(\bar{\bm x}_{j}) \hat{\phi}(\bar{\mf x}_{j}),
    \end{equation}
    where $f_j$ is the smearing function for the $j$th detector,  $\hat{\mu}_j(\tau_{j})$ is its monopole moment and $\lambda_j$ the coupling strength. 
    
    In Section \ref{timeOrderSection} we obtained results for a general interaction Hamiltonian weight\textcolor{black}{. W}e can now apply those results to the multiple detectors case where the Hamiltonian weight is 
    \begin{equation}\label{hIMultiple}
        \hat{h}_I(\mathsf{x}) = \sum_{j=1}^{N} \hat{h}_{I,j}(\mathsf{x}).
    \end{equation}
   The time-evolution calculations can get quite complicated if the detectors are in different states of motion. This is because  to obtain the total Hamiltonian or the time evolution operator, in general, we need to recast all the summands in  \eqref{hIMultiple} in terms of a common set of coordinates different from at least some of the detector's proper frame. Notice, however, that in the case of pointlike detectors, and therefore with Dirac deltas as smearings $f_j(\bar{\bm x}_{j})$, the Hamiltonian weight  from Equation \eqref{hIMultiple} commutes with itself at spacelike separated points. This is due to the fact that the different monopole moment operators act in different Hilbert spaces and the field operator is assumed to satisfy the axiom of microcausality. Therefore, we conclude that for a system of pointlike detectors, the time evolution operator can be written as
    \begin{equation}
        \hat{\mathcal{U}} = \mathcal{T}_{}\exp\left(-\ii \int_\mathcal{M}\!\!\! \dd\mathcal{V} \hat{h}_I(\mf x)\right),
    \end{equation}
    with no necessity to explicitly indicate with respect to which time parameter the ordering happens. In other words, as anticipated in previous sections, the formalism for (an arbitrary number of) pointlike UDW detectors is fully covariant. 
    
    Same as in the single-detector case, violations of covariance will appear when smeared detectors are considered. 
    \color{black} In lieu of full covariance for one detector, one may be tempted to privilege the time ordering with respect to the proper time of the detector's center of mass due to the fact that interaction is prescribed in the Fermi-Walker reference frame of the detector's centre of mass. However, when multiple detectors are considered we are mixing different Hamiltonian weights prescribed with respect to different Fermi-Walker frames. The results therefore would be different if we time order the full interaction with respect to any of the many proper time parameters involved in the many-detector problem. In plain words, should we time-order the global $\hat{\mathcal{U}}$ with respect to Alice's detector's proper time? or Bob's? Or Charles's? Or none? Each prescription would yield quantitatively different predictions. Obviously this is a problem: there is no unique way of writing the time evolution operator for a system of $N$ smeared particle detectors.
    
    Since the UDW model is an effective model, we do not necessarily expect that it is fully covariant, as all fundamental models must be. The breakdown of covariance just responds to the usage of the effective model beyond the regimes in which it can be used to properly model the (covariant) physical reality. With this in mind, we will show that there are physically reasonable regimes where the covariance breakdown can be minimized and provide approximate results that can be used to define limits of validity of the theory.
    
   To quantify the dependence of the predictions of the smeared UDW model on the coordinate system used, let us first consider an arbitrary choice of time parameter $s$ associated to a foliation $\mathcal{E}_s$, yielding the following time evolution operator \color{black} 
    \begin{equation}
         \hat{\mathcal{U}}_s = \mathcal{T}_{s}\exp\left(-\ii \int_\mathcal{M}\!\!\! \dd\mathcal{V} \hat{h}_I(\mf x)\right).
    \end{equation}
    
    In the case of multiple detectors, we can adapt the calculation done in Section \ref{CovOneDetector} to prove a similar result: choosing an initial detectors' state that commutes with their free Hamiltonian cancels the violations of covariance at $\mathcal{O}(\lambda^2)$. Moreover, if the field state is Gaussian with a zero one-point function the difference in the predictions for $\hat{\mathcal{U}}$ with respect to different time parameters is cancelled also at $\mathcal{O}(\lambda^3)$.
     
     Furthermore, for arbitrary states, the offending deviation can be calculated at leading order from equation \eqref{U2correction} by plugging in the Hamiltonian weight \eqref{hIMultiple}. Let us consider the second order term in the Dyson expansion prescribed with respect to two notions of time ordering, $t$ and $s$ that do not necessarily agree. We then obtain two time evolution operators, $\hat{\mathcal{U}}_t$ and $\hat{\mathcal{U}}_s$ with their associated second order terms being $\hat{\mathcal{U}}_t^{(2)}$ and $\hat{\mathcal{U}}_s^{(2)}$. Recalling that in the region $S_\leq$ the field operators commute, and so do the monopole operators associated to different detectors, we have  that $\big[\hat{h}_{I,i}(\mathsf{x}),\hat{h}_{I,j}(\mathsf{x}')\big] = 0$ for $i\neq j$ in $S_\leq$, and Eq. \eqref{U2correction} yields
    \begin{equation}\label{deviation}
        \hat{\mathcal{U}}_t^{(2)} - \hat{\mathcal{U}}_s^{(2)} = - \sum_{i = 1}^N\int_{S_\leq} \!\!\!\dd \mathcal{V} \dd \mathcal{V}' \comm{\hat{h}_{I,i}(\mathsf{x})}{\hat{h}_{I,i}(\mathsf{x}')}\theta\big(t' - t\big).
    \end{equation}
    This gives us
    \begin{equation}
         \hat{\mathcal{U}}_t^{(2)} - \hat{\mathcal{U}}_s^{(2)} = \lambda^2 \sum_{i = 1}^N\hat{\sigma}_{z,i} \hat{E}_i,
         \end{equation}
    where $\hat{E}_i$ corresponds exactly to the $\hat{E}$ defined in \eqref{defE} for each detector. If the system starts in an uncorrelated state of the form
    \begin{equation}
        \hat{\rho}_0 = \left(\bigotimes_{i = 1}^N\hat{\rho}_{0, i}\right)\otimes\hat{\rho}_\phi,
    \end{equation}
    the same procedure outlined in section \ref{CovOneDetector} leads to two different density operators for the detector part of the system, $\hat{\rho}^s_{\textrm{d}}$ associated to time evolution with respect to the parameter $s$, and $\hat{\rho}^t_{\textrm{d}}$ associated to the parameter $t$. Their difference will then be given by
    \begin{equation}\label{last}
        \hat{\rho}_\textrm{d}^t-\hat{\rho}^s_\textrm{d} =  \lambda^2\sum_{i=1}^N\left(\bigotimes_{j \neq i}\hat{\rho}_{0, j}\right)\comm{\hat{\sigma}_{z,i}}{\hat{\rho}_{0,i}}\Tr \hat{\rho}_\phi \hat{E}_i+\mathcal{O}(\lambda^3).
    \end{equation}
    It is therefore clear that if all detectors start in a product state, with the state of each detector being a statistical mixture of eigenstates of the respective free Hamiltonian, the deviation up to second order in the coupling vanishes\color{black}, allowing the model to be used within these regimes. 
    

    This means that although there is no unique non-perturbative way of writing a given time evolution operator for smeared detectors, we do not see any difference in predictions for different time ordering at leading order in the coupling. It is important to remark that the standard results obtained from techniques and setups that are dependent on multiple UDW detectors, such as entanglement harvesting and quantum energy teleportation, are dominated by second order dynamics and often use initial states for which the second order violation cancels. In all those cases there is no violation of covariance in the final result.
    \color{black}

    Moreover, same as in the case of a single detector, the violations of covariance scale with the size of the detectors as it can be seen from the definition of the $\hat E_i$ operators. This means that the violation of covariance can be made small under the following three conditions: 1) the relative motion of the detectors with respect to the frame in which we are computing $\hat{\mathcal{U}}$ is not extreme, 2) the curvature around the detectors is also not extreme, and 3) the predictions are going to be considered for times much longer than the light-crossing time of the lenghtscale of each of the detectors in their respective proper frames. In those cases, making the detector smaller suppresses the covariance violations very fast. For atomic-sized detectors one would expect these three assumptions to hold even for regimes where the Unruh effect is detectable, paralleling the discussion about orders of magnitude where these effects are relevant found in \cite{us}. We will illustrate this decay of the violations of covariance with an example in the next section.
    
    \section{Example: Smeared Inertial detector in 
    Flat spacetime}\label{ExampleFlatSpacetime}
    
    Even the simplest possible dynamics for the detector and field---inertial motion in flat spacetimes---already suffers from the covariance violation studied in this paper. That is, the UDW model for an inertial detector (of center of mass proper time $\tau$) moving with respect to the frame used for the quantization of a scalar quantum field $(t,\bm{x})$ (that we call the lab frame) still yields different predictions if the time ordering is taken with respect to $\tau$ or $t$. Evaluating Equation \eqref{defE} explicitly for this simple case will provide intuition on the scales that play a role in determining the regimes where the breaking of covariance can be neglected.
    
    Without loss of generality, we can take the detector's centre of mass to be moving in the $x$ direction, with positive speed $v$ relative to the lab frame. We make the choice of Fermi-Walker coordinates for the detector $(\tau, \bar{x}, \bar{\bm{x}}_\perp)$, so that $\bar{\bm{x}}_\perp$   comprises the coordinates in the spatial directions that are orthogonal to the detector's velocity. The lab time $\Delta t$ elapsed between two events with coordinates $(\tau, \bar{x}, \bar{\bm{x}}_\perp)$ and $(\tau', \bar{x}', \bar{\bm{x}}'_\perp)$ is simply given by a Lorentz transformation:
    \begin{equation}
    \begin{gathered}
        \Delta t = \gamma\left(\Delta\tau + v\Delta   \bar{x}\right),\\
        \gamma \equiv \dfrac{1}{\sqrt{1 - v^2}}, 
    \end{gathered}
    \end{equation}
where $\Delta\tau = \tau - \tau'$, $\Delta\bar{x} = \bar{x} - \bar{x}'$. Time ordering is different in the two frames only for events in the region $S_\leq$, since in that region $\Delta\tau > 0$ and $\Delta t < 0$. This happens when
\begin{equation}
    \Delta\bar{x} < -\dfrac{\Delta \tau}{v}.
\end{equation}
Therefore, in this case, the region $S_\leq$ can be written in the Fermi normal coordinates of the detector as the points $(\mathsf{\bar{x}}, \mathsf{\bar{x}}')$ parametrized by 
\begin{equation}\label{parametrization}
    \begin{gathered}
     \bar{\mf{x}} = (\tau, \bar{x}, \bar{\bm{x}}_\perp), \\
     \bar{\mf{x}}' = (\tau - \sigma, \bar{x} - \xi, \bar{\bm{x}}_\perp')
    \end{gathered}
\end{equation}
with $\sigma > 0$, $\xi < -\sigma/v$, and $\bar{\bm{x}}_\perp, \bar{\bm{x}}_\perp'$ arbitrary. 


One primary consistency check for our previous claims  is to see that Eq. \eqref{defE} vanishes when we set the smearing function to be \mbox{$f(\bar{x}, \bar{\bm{x}}_\perp) = \delta(\bar{x})\delta^{(n-1)}(\bar{\bm{x}}_\perp)$}, which would correspond to the case of a pointlike detector. With this choice of smearing and the parametrization of the region $S_\leq$ according to Eq.~\eqref{parametrization}, the integrals  over $\bar{x}, \bar{\bm{x}}_\perp, \bar{\bm{x}}'_\perp$ in Eq.~\eqref{defE} can be trivially computed in the case of a  stationary state of the field, so that we are left with
\begin{align}\label{inertialerror}
    \Tr \hat{\rho}_\phi \hat{E} =&\, 2\ii\int_{0}^{\infty}\!\!\!\!\dd\sigma\int_{-\infty}^{-\sigma/v}\!\!\!\!\!\!\!\!\!\!\!\dd\xi\,\, \delta(\xi)\sin(\Omega\sigma)\mathcal{W}_{\hat{\rho}_\phi}(\xi,\sigma)\nonumber\\
    &\times\int_{\mathbb{R}}\!\dd\tau \chi(\tau) \chi(\tau - \sigma),
\end{align}
where 
\begin{equation}
    \mathcal{W}_{\hat{\rho}_\phi}(\xi,\sigma)\coloneqq \ev{\hat{\phi}(\mf x(0,0,\bm 0)) \hat{\phi}(\mf x(-\sigma, -\xi, \bm 0))}_{\hat\rho_\phi}
\end{equation} 
is obtained from the field's Wightman function $\langle\hat\phi(\mf{x})\hat\phi(\mf{x}')\rangle$ assuming stationarity and after carrying out all the spatial integrals but $\xi$ using the delta smearing.

Since the domain of integration in $\xi$ never crosses the origin, the integral in Eq. \eqref{inertialerror} yields zero. This is consistent with what we showed in Section \ref{CovOneDetector}: pointlike detectors do not introduce any covariance problems.


We now compute explicitly the deviation from predictions between  time-ordering with detector's proper time and an arbitrary inertial frame. For concreteness, let us consider the vacuum state of the field in three spatial dimensions. The vacuum Wightman function of a massless scalar field  evaluated between spacelike points is given by:
\begin{equation}\label{wightman}
    \bra{0}\!\hat{\phi}(\mathsf{x})\hat{\phi}(\mathsf{x}')\!\ket{0} = \dfrac{2}{{(2\pi)^2}}\dfrac{1}{\abs{\Delta \mathsf{x}}^2},
\end{equation}
where $\abs{\Delta\mathsf{x}}^2 = \eta_{\mu\nu}(\Delta\mathsf{x})^\mu(\Delta\mathsf{x})^\nu$ is the invariant spacetime interval between the events $\mathsf{x}$ and $\mathsf{x}'$. In the coordinates associated to the frame of the detector, $\abs{\Delta \mf x}^2$ can be written as
\begin{equation}\label{deltaX}
     \abs{\Delta\mathsf{x}}^2 = -\sigma^2 + \xi^2 + \abs{\bar{\bm{x}}_\perp - \bar{\bm{x}}'_\perp}^2.
\end{equation}

We consider a Gaussian switching function $\chi(\tau)$ with timescale $T$ and a Gaussian smearing function $f(\bm{\bar{x}})$ with length scale $\ell$ respectively:
\vspace*{-8pt}
\begin{align}
    \chi(\tau) &= \dfrac{1}{\sqrt{2\pi}}\exp\left(-\dfrac{\tau^2}{2T^2}\right),
\end{align}
\vspace*{-18pt}
\begin{align}
    f(\bar{\bm{x}}) &= \dfrac{1}{\sqrt{(2\pi)^3}\ell^3}\exp\left(-\dfrac{\abs{\bar{\bm{x}}}^2}{2\ell^2}\right).
\end{align}
With these choices for switching and smearing, the integrals over $\tau$ and $\bar{x}$ in \eqref{defE} can be computed in closed form. The integrals in the perpendicular directions can be evaluated by changing variables from $\bar{\bm{x}}_\perp, \bar{\bm{x}}_\perp'$ to $\bm{r} = \bar{\bm{x}}_\perp - \bar{\bm{x}}_\perp'$ and $\bm{R} = \bar{\bm{x}}_\perp + \bar{\bm{x}}_\perp'$. By doing so, Eq. \eqref{defE} takes the following form
\begin{widetext}
\begin{align}
    \Tr \hat{\rho}_\phi \hat{E} =& \dfrac{4\ii}{{(2\pi)^2}} \int_{\mathbb{R}^2}\!\!\dd^{2}\bar{\bm x}_\perp\int_{\mathbb{R}^2}\!\!\dd^{2}\bar{\bm x}_\perp'\int_{\mathbb{R}^2}\!\!\!\dd\tau\dd \bar{x}\int_{0}^{\infty}\!\!\!\dd\sigma\int_{-\infty}^{-\sigma/v}\!\!\!\dd \xi \,\dfrac{\chi(\tau)\chi(\tau - \sigma)\sin(\Omega\sigma)f(\bar{x}, \bar{\bm{x}}_\perp)f(\bar{x} - \xi, \bar{\bm{x}}_\perp')}{\left(-\sigma^2 + \xi^2 + \abs{\bar{\bm{x}}_\perp - \bar{\bm{x}}'_\perp}^2\right)}\nonumber\\
    =& \dfrac{\ii T}{{2 \pi^2}\ell^3} \int_{0}^{\infty}\dd \sigma \int_{-\infty}^{-\sigma/v}\dd \xi\, e^{-\sigma^2/4T^2}e^{-\xi^2/4
    \ell^2}\sin(\Omega\sigma)\int_{\mathbb{R}^2}\dd^2r\dfrac{e^{-\abs{\bm{r}}^2/4\ell^2}}{\xi^2 - \sigma^2 + \abs{\bm{r}}^2} \nonumber\\
     =&  \dfrac{\ii T}{{\pi}\ell^3} \int_{0}^{\infty}\dd \sigma \int_{-\infty}^{-\sigma/v}\dd \xi\, e^{-\sigma^2/4T^2}e^{-\xi^2/4
    \ell^2}\sin(\Omega\sigma)\int_{0}^{\infty}\dd r\dfrac{re^{-r^2/4\ell^2}}{\xi^2 - \sigma^2 + r^2} \nonumber\\
    =& \dfrac{\ii T}{2{\pi}\ell^3}\int_{0}^{\infty}\dd \sigma \int_{-\infty}^{-\sigma/v}\dd \xi\,\exp\left(-\sigma^2\left(\dfrac{1}{4T^2} + \dfrac{1}{4\ell^2}\right)\right)\sin(\Omega\sigma)\,\operatorname{Ei}\left(\tfrac{-\xi^2 + \sigma^2}{4\ell^2}\right) \label{error1}\\
     =& \dfrac{\ii }{{\pi}}\left(\dfrac{T}{\ell}\right)^3 v\int_{0}^{\infty}\dd s \int_{-\infty}^{-s}\dd \zeta\,\exp\left(-\dfrac{s^2 v^2}{4}\left(1 + \dfrac{T^2}{\ell^2}\right)\right)\sin(\Omega T v s)\,\operatorname{Ei}\left((-\zeta^2 + s^2v^2)\dfrac{T^2}{4\ell^2}\right) \label{error2},
    \end{align}
\end{widetext}
    where $\operatorname{Ei}(x)$ is the exponential integral function~\cite{database}, and we get equation \eqref{error2} from \eqref{error1} by performing the change of variables $s = \sigma/vT$, $\zeta = \xi/T$. Analysis on \eqref{error2} shows that for fixed $v$, the violation of covariance computed above goes to zero as the duration of the interaction $T$ becomes much longer than the light-crossing time of the detector $\ell$. Numerical results show that for values of $T/\ell\gtrapprox 10^3$ the error becomes negligible for speeds below $v\leq0.9$. As a summary, in the limit of $T/\ell \rightarrow \infty$, the whole integrand in equation \eqref{error2} vanishes, and therefore so does the covariance breaking term as expected from the discussion in previous sections.

    \section{Conclusion}
        
    We have studied the breakdown of covariance that the time-ordering operation introduces in smeared particle detector models (such as the UDW model) used in QFT in general spacetimes.
    
    We have first shown how for pointlike detectors, the time-ordering operation does not introduce any coordinate dependence: all predictions of properly prescribed pointlike UDW detectors are covariant. Namely, we have explicitly shown how, for the predictions of a system of $N$ pointlike particle detectors on arbitrary trajectories in curved spacetimes, all possible choices of time-ordering are equivalent. We highlighted that all predictions are covariant even when the multiple pointlike detectors are relatively spacelike separated. This can be traced back to the fact that a) pointlike detectors only see the field along timelike trajectories---so the time ordering of the events making up each detector's worldline is unambiguous---and b) the individual Hamiltonian densities coupling each detector to the field mutually commute when the detectors are spacelike separated. 
   
    In contrast, we have shown that, for smeared detectors, the fact that the detectors couple to the field at multiple spacelike separated points introduces a break of covariance in time-ordering. This is problematic because different choices of time-ordering parameter can, in principle, yield radically different predictions. This is aggravated for systems of many detectors in arbitrary states of motion since there is no physical reason in those setups to privilege one particular notion of time order.
    
    With this in mind, we explicitly evaluated the magnitude of this break of covariance and concluded that if a detector starts in a statistical mixture of eigenstates of its free Hamiltonian (such as ground, excited or thermal state), the deviations from a fully covariant prediction are of third order in the detector's coupling strength (and in most cases even fourth order), hence subleading for many interesting phenomena (e.g., the thermal response of detectors in the Unruh and Hawking effects~\cite{Unruh1976,Sciama1977,Unruh-Wald,Takagi,Louko} and typical scenarios of entanglement harvesting \cite{Valentini1991,Reznik1,reznik2,Retzker2005,RalphOlson1,RalphOlson2,Nick,Cosmo,Salton:2014jaa,Pozas-Kerstjens:2015}). Furthermore, in the cases where the breakdown of covariance is of leading order, we have argued that it is of the same magnitude as the causality violation already introduced by the mere fact of smearing a detector degree of freedom~\cite{martin-martinez2015}, and showed that these deviations from covariance are suppressed with the smearing length scales. Analogously to the discussion in  \cite{us},the difference between predictions in different coordinates can be negligible in scenarios where the duration of the interaction is much longer than the light-crossing time of the detector’s smearing length scale in all the detectors’ center of mass frames and in the coordinate frame used to perform calculations. We have also shown a particular example of this in flat spacetime. 
    
    The analysis on this paper quantifies the coordinate dependence of predictions for particle detector models in a very general setting as a function of the initial states, the shape and state of motion of the detectors, and the geometry of the spacetime they move in. Thus, these results establish the limits of validity of smeared particle detector models to covariantly extract information from a quantum field. 
    
        
    \section{Acknowledgements}
    The authors thank Luis J. Garay, Jonas Neuser and Erickson Tjoa for insightful discussions. E.M-M acknowledges the support of the NSERC Discovery program as well as his Ontario Early Researcher Award. T.R.P. and B.S.L.T. thank IFT-UNESP/ICTP-SAIFR and CAPES for partial financial support. Research at Perimeter Institute is supported in part by the Government of Canada through the Department of Innovation, Science and Economic Development Canada and by the Province of Ontario through the Ministry of Colleges and Universities.

\twocolumngrid
\bibliography{references}

\begin{thebibliography}{35}%
\makeatletter
\providecommand \@ifxundefined [1]{%
 \@ifx{#1\undefined}
}%
\providecommand \@ifnum [1]{%
 \ifnum #1\expandafter \@firstoftwo
 \else \expandafter \@secondoftwo
 \fi
}%
\providecommand \@ifx [1]{%
 \ifx #1\expandafter \@firstoftwo
 \else \expandafter \@secondoftwo
 \fi
}%
\providecommand \natexlab [1]{#1}%
\providecommand \enquote  [1]{``#1''}%
\providecommand \bibnamefont  [1]{#1}%
\providecommand \bibfnamefont [1]{#1}%
\providecommand \citenamefont [1]{#1}%
\providecommand \href@noop [0]{\@secondoftwo}%
\providecommand \href [0]{\begingroup \@sanitize@url \@href}%
\providecommand \@href[1]{\@@startlink{#1}\@@href}%
\providecommand \@@href[1]{\endgroup#1\@@endlink}%
\providecommand \@sanitize@url [0]{\catcode `\\12\catcode `\$12\catcode
  `\&12\catcode `\#12\catcode `\^12\catcode `\_12\catcode `\%12\relax}%
\providecommand \@@startlink[1]{}%
\providecommand \@@endlink[0]{}%
\providecommand \url  [0]{\begingroup\@sanitize@url \@url }%
\providecommand \@url [1]{\endgroup\@href {#1}{\urlprefix }}%
\providecommand \urlprefix  [0]{URL }%
\providecommand \Eprint [0]{\href }%
\providecommand \doibase [0]{http://dx.doi.org/}%
\providecommand \selectlanguage [0]{\@gobble}%
\providecommand \bibinfo  [0]{\@secondoftwo}%
\providecommand \bibfield  [0]{\@secondoftwo}%
\providecommand \translation [1]{[#1]}%
\providecommand \BibitemOpen [0]{}%
\providecommand \bibitemStop [0]{}%
\providecommand \bibitemNoStop [0]{.\EOS\space}%
\providecommand \EOS [0]{\spacefactor3000\relax}%
\providecommand \BibitemShut  [1]{\csname bibitem#1\endcsname}%
\let\auto@bib@innerbib\@empty
\bibitem [{\citenamefont {Unruh}(1976)}]{Unruh1976}%
  \BibitemOpen
  \bibfield  {author} {\bibinfo {author} {\bibfnamefont {W.~G.}\ \bibnamefont
  {Unruh}},\ }\href {\doibase 10.1103/PhysRevD.14.870} {\bibfield  {journal}
  {\bibinfo  {journal} {Phys. Rev. D}\ }\textbf {\bibinfo {volume} {14}},\
  \bibinfo {pages} {870} (\bibinfo {year} {1976})}\BibitemShut {NoStop}%
\bibitem [{\citenamefont {Unruh}\ and\ \citenamefont
  {Wald}(1984)}]{Unruh-Wald}%
  \BibitemOpen
  \bibfield  {author} {\bibinfo {author} {\bibfnamefont {W.~G.}\ \bibnamefont
  {Unruh}}\ and\ \bibinfo {author} {\bibfnamefont {R.~M.}\ \bibnamefont
  {Wald}},\ }\href {\doibase 10.1103/PhysRevD.29.1047} {\bibfield  {journal}
  {\bibinfo  {journal} {Phys. Rev. D}\ }\textbf {\bibinfo {volume} {29}},\
  \bibinfo {pages} {1047} (\bibinfo {year} {1984})}\BibitemShut {NoStop}%
\bibitem [{\citenamefont {DeWitt}(1980)}]{DeWitt}%
  \BibitemOpen
  \bibfield  {author} {\bibinfo {author} {\bibfnamefont {B.}~\bibnamefont
  {DeWitt}},\ }\href@noop {} {\emph {\bibinfo {title} {General Relativity; an
  Einstein Centenary Survey}}}\ (\bibinfo  {publisher} {Cambridge University
  Press},\ \bibinfo {address} {Cambridge, UK},\ \bibinfo {year}
  {1980})\BibitemShut {NoStop}%
\bibitem [{\citenamefont {Sorkin}(1993)}]{Sorkin}%
  \BibitemOpen
  \bibfield  {author} {\bibinfo {author} {\bibfnamefont {R.~D.}\ \bibnamefont
  {Sorkin}},\ }\href@noop {} {\enquote {\bibinfo {title} {Impossible
  measurements on quantum fields},}\ } (\bibinfo {year} {1993}),\ \Eprint
  {http://arxiv.org/abs/gr-qc/9302018} {arXiv:gr-qc/9302018 [gr-qc]}
  \BibitemShut {NoStop}%
\bibitem [{\citenamefont {Benincasa}\ \emph {et~al.}(2014)\citenamefont
  {Benincasa}, \citenamefont {Borsten}, \citenamefont {Buck},\ and\
  \citenamefont {Dowker}}]{Benincasa_2014}%
  \BibitemOpen
  \bibfield  {author} {\bibinfo {author} {\bibfnamefont {D.~M.~T.}\
  \bibnamefont {Benincasa}}, \bibinfo {author} {\bibfnamefont {L.}~\bibnamefont
  {Borsten}}, \bibinfo {author} {\bibfnamefont {M.}~\bibnamefont {Buck}}, \
  and\ \bibinfo {author} {\bibfnamefont {F.}~\bibnamefont {Dowker}},\ }\href
  {\doibase 10.1088/0264-9381/31/7/075007} {\bibfield  {journal} {\bibinfo
  {journal} {Class. Quantum Gravity}\ }\textbf {\bibinfo {volume} {31}},\
  \bibinfo {pages} {075007} (\bibinfo {year} {2014})}\BibitemShut {NoStop}%
\bibitem [{\citenamefont {Bostelmann}\ \emph {et~al.}(2020)\citenamefont
  {Bostelmann}, \citenamefont {Fewster},\ and\ \citenamefont
  {Ruep}}]{fewster3}%
  \BibitemOpen
  \bibfield  {author} {\bibinfo {author} {\bibfnamefont {H.}~\bibnamefont
  {Bostelmann}}, \bibinfo {author} {\bibfnamefont {C.~J.}\ \bibnamefont
  {Fewster}}, \ and\ \bibinfo {author} {\bibfnamefont {M.~H.}\ \bibnamefont
  {Ruep}},\ }\href@noop {} {\enquote {\bibinfo {title} {Impossible measurements
  require impossible apparatus},}\ } (\bibinfo {year} {2020}),\ \Eprint
  {http://arxiv.org/abs/2003.04660} {arXiv:2003.04660 [quant-ph]} \BibitemShut
  {NoStop}%
\bibitem [{\citenamefont {Borsten}\ \emph {et~al.}(2019)\citenamefont
  {Borsten}, \citenamefont {Jubb},\ and\ \citenamefont {Kells}}]{borsten}%
  \BibitemOpen
  \bibfield  {author} {\bibinfo {author} {\bibfnamefont {L.}~\bibnamefont
  {Borsten}}, \bibinfo {author} {\bibfnamefont {I.}~\bibnamefont {Jubb}}, \
  and\ \bibinfo {author} {\bibfnamefont {G.}~\bibnamefont {Kells}},\
  }\href@noop {} {\enquote {\bibinfo {title} {Impossible measurements
  revisited},}\ } (\bibinfo {year} {2019}),\ \Eprint
  {http://arxiv.org/abs/1912.06141} {arXiv:1912.06141 [quant-ph]} \BibitemShut
  {NoStop}%
\bibitem [{\citenamefont {Candelas}\ and\ \citenamefont
  {Sciama}(1977)}]{Sciama1977}%
  \BibitemOpen
  \bibfield  {author} {\bibinfo {author} {\bibfnamefont {P.}~\bibnamefont
  {Candelas}}\ and\ \bibinfo {author} {\bibfnamefont {D.~W.}\ \bibnamefont
  {Sciama}},\ }\href {\doibase 10.1103/PhysRevLett.38.1372} {\bibfield
  {journal} {\bibinfo  {journal} {Phys. Rev. Lett.}\ }\textbf {\bibinfo
  {volume} {38}},\ \bibinfo {pages} {1372} (\bibinfo {year}
  {1977})}\BibitemShut {NoStop}%
\bibitem [{\citenamefont {Takagi}(1986)}]{Takagi}%
  \BibitemOpen
  \bibfield  {author} {\bibinfo {author} {\bibfnamefont {S.}~\bibnamefont
  {Takagi}},\ }\href {\doibase 10.1143/PTP.88.1} {\bibfield  {journal}
  {\bibinfo  {journal} {Prog. Theor. Phys. Supp.}\ }\textbf {\bibinfo {volume}
  {88}},\ \bibinfo {pages} {1} (\bibinfo {year} {1986})}\BibitemShut {NoStop}%
\bibitem [{\citenamefont {Hodgkinson}\ \emph {et~al.}(2014)\citenamefont
  {Hodgkinson}, \citenamefont {Louko},\ and\ \citenamefont {Ottewill}}]{Louko}%
  \BibitemOpen
  \bibfield  {author} {\bibinfo {author} {\bibfnamefont {L.}~\bibnamefont
  {Hodgkinson}}, \bibinfo {author} {\bibfnamefont {J.}~\bibnamefont {Louko}}, \
  and\ \bibinfo {author} {\bibfnamefont {A.~C.}\ \bibnamefont {Ottewill}},\
  }\href {\doibase 10.1103/PhysRevD.89.104002} {\bibfield  {journal} {\bibinfo
  {journal} {Phys. Rev. D}\ }\textbf {\bibinfo {volume} {89}},\ \bibinfo
  {pages} {104002} (\bibinfo {year} {2014})}\BibitemShut {NoStop}%
\bibitem [{\citenamefont {Pozas-Kerstjens}\ and\ \citenamefont
  {Mart\'{i}n-Mart\'{i}nez}(2016)}]{Pozas2016}%
  \BibitemOpen
  \bibfield  {author} {\bibinfo {author} {\bibfnamefont {A.}~\bibnamefont
  {Pozas-Kerstjens}}\ and\ \bibinfo {author} {\bibfnamefont {E.}~\bibnamefont
  {Mart\'{i}n-Mart\'{i}nez}},\ }\href {\doibase 10.1103/PhysRevD.94.064074}
  {\bibfield  {journal} {\bibinfo  {journal} {Phys. Rev. D}\ }\textbf {\bibinfo
  {volume} {94}},\ \bibinfo {pages} {064074} (\bibinfo {year}
  {2016})}\BibitemShut {NoStop}%
\bibitem [{\citenamefont {Mart\'{i}n-Mart\'{i}nez}\ and\ \citenamefont
  {Rodriguez-Lopez}(2018)}]{eduardo}%
  \BibitemOpen
  \bibfield  {author} {\bibinfo {author} {\bibfnamefont {E.}~\bibnamefont
  {Mart\'{i}n-Mart\'{i}nez}}\ and\ \bibinfo {author} {\bibfnamefont
  {P.}~\bibnamefont {Rodriguez-Lopez}},\ }\href {\doibase
  10.1103/PhysRevD.97.105026} {\bibfield  {journal} {\bibinfo  {journal} {Phys.
  Rev. D}\ }\textbf {\bibinfo {volume} {97}},\ \bibinfo {pages} {105026}
  (\bibinfo {year} {2018})}\BibitemShut {NoStop}%
\bibitem [{\citenamefont {Schlicht}(2004)}]{Schlicht}%
  \BibitemOpen
  \bibfield  {author} {\bibinfo {author} {\bibfnamefont {S.}~\bibnamefont
  {Schlicht}},\ }\href {\doibase 10.1088/0264-9381/21/19/011} {\bibfield
  {journal} {\bibinfo  {journal} {Class. Quantum Gravity}\ }\textbf {\bibinfo
  {volume} {21}},\ \bibinfo {pages} {4647} (\bibinfo {year}
  {2004})}\BibitemShut {NoStop}%
\bibitem [{\citenamefont {Louko}\ and\ \citenamefont {Satz}(2006)}]{Jorma}%
  \BibitemOpen
  \bibfield  {author} {\bibinfo {author} {\bibfnamefont {J.}~\bibnamefont
  {Louko}}\ and\ \bibinfo {author} {\bibfnamefont {A.}~\bibnamefont {Satz}},\
  }\href {\doibase 10.1088/0264-9381/23/22/015} {\bibfield  {journal} {\bibinfo
   {journal} {Class. Quantum Gravity}\ }\textbf {\bibinfo {volume} {23}},\
  \bibinfo {pages} {6321} (\bibinfo {year} {2006})}\BibitemShut {NoStop}%
\bibitem [{\citenamefont {Fewster}\ and\ \citenamefont
  {Rejzner}(2019)}]{kasia}%
  \BibitemOpen
  \bibfield  {author} {\bibinfo {author} {\bibfnamefont {C.~J.}\ \bibnamefont
  {Fewster}}\ and\ \bibinfo {author} {\bibfnamefont {K.}~\bibnamefont
  {Rejzner}},\ }\href@noop {} {\enquote {\bibinfo {title} {Algebraic quantum
  field theory -- an introduction},}\ } (\bibinfo {year} {2019}),\ \Eprint
  {http://arxiv.org/abs/1904.04051} {arXiv:1904.04051 [hep-th]} \BibitemShut
  {NoStop}%
\bibitem [{\citenamefont {Scully}\ and\ \citenamefont
  {Zubairy}(1997)}]{ScullyBook}%
  \BibitemOpen
  \bibfield  {author} {\bibinfo {author} {\bibfnamefont {M.~O.}\ \bibnamefont
  {Scully}}\ and\ \bibinfo {author} {\bibfnamefont {M.~S.}\ \bibnamefont
  {Zubairy}},\ }\href {\doibase 10.1017/CBO9780511813993} {\emph {\bibinfo
  {title} {Quantum Optics}}}\ (\bibinfo  {publisher} {Cambridge University
  Press},\ \bibinfo {year} {1997})\BibitemShut {NoStop}%
\bibitem [{\citenamefont
  {Mart\'{\i}n-Mart\'{\i}nez}(2015)}]{martin-martinez2015}%
  \BibitemOpen
  \bibfield  {author} {\bibinfo {author} {\bibfnamefont {E.}~\bibnamefont
  {Mart\'{\i}n-Mart\'{\i}nez}},\ }\href {\doibase 10.1103/PhysRevD.92.104019}
  {\bibfield  {journal} {\bibinfo  {journal} {Phys. Rev. D}\ }\textbf {\bibinfo
  {volume} {92}},\ \bibinfo {pages} {104019} (\bibinfo {year}
  {2015})}\BibitemShut {NoStop}%
\bibitem [{\citenamefont {Mart\'{\i}n-Mart\'{\i}nez}\ \emph
  {et~al.}(2020)\citenamefont {Mart\'{\i}n-Mart\'{\i}nez}, \citenamefont
  {Perche},\ and\ \citenamefont {de~S.~L.~Torres}}]{us}%
  \BibitemOpen
  \bibfield  {author} {\bibinfo {author} {\bibfnamefont {E.}~\bibnamefont
  {Mart\'{\i}n-Mart\'{\i}nez}}, \bibinfo {author} {\bibfnamefont {T.~R.}\
  \bibnamefont {Perche}}, \ and\ \bibinfo {author} {\bibfnamefont
  {B.}~\bibnamefont {de~S.~L.~Torres}},\ }\href {\doibase
  10.1103/PhysRevD.101.045017} {\bibfield  {journal} {\bibinfo  {journal}
  {Phys. Rev. D}\ }\textbf {\bibinfo {volume} {101}},\ \bibinfo {pages}
  {045017} (\bibinfo {year} {2020})}\BibitemShut {NoStop}%
\bibitem [{\citenamefont {Valentini}(1991)}]{Valentini1991}%
  \BibitemOpen
  \bibfield  {author} {\bibinfo {author} {\bibfnamefont {A.}~\bibnamefont
  {Valentini}},\ }\href {\doibase
  http://dx.doi.org/10.1016/0375-9601(91)90952-5} {\bibfield  {journal}
  {\bibinfo  {journal} {Phys. Lett. A}\ }\textbf {\bibinfo {volume} {153}},\
  \bibinfo {pages} {321 } (\bibinfo {year} {1991})}\BibitemShut {NoStop}%
\bibitem [{\citenamefont {Reznik}\ \emph {et~al.}(2005)\citenamefont {Reznik},
  \citenamefont {Retzker},\ and\ \citenamefont {Silman}}]{Reznik1}%
  \BibitemOpen
  \bibfield  {author} {\bibinfo {author} {\bibfnamefont {B.}~\bibnamefont
  {Reznik}}, \bibinfo {author} {\bibfnamefont {A.}~\bibnamefont {Retzker}}, \
  and\ \bibinfo {author} {\bibfnamefont {J.}~\bibnamefont {Silman}},\ }\href
  {http://link.aps.org/abstract/PRA/v71/e042104} {\bibfield  {journal}
  {\bibinfo  {journal} {Phys. Rev. A}\ }\textbf {\bibinfo {volume} {71}},\
  \bibinfo {eid} {042104} (\bibinfo {year} {2005})}\BibitemShut {NoStop}%
\bibitem [{\citenamefont {Silman}\ and\ \citenamefont
  {Reznik}(2007)}]{reznik2}%
  \BibitemOpen
  \bibfield  {author} {\bibinfo {author} {\bibfnamefont {J.}~\bibnamefont
  {Silman}}\ and\ \bibinfo {author} {\bibfnamefont {B.}~\bibnamefont
  {Reznik}},\ }\href@noop {} {\bibfield  {journal} {\bibinfo  {journal} {Phys.
  Rev. A}\ }\textbf {\bibinfo {volume} {75}},\ \bibinfo {pages} {052307}
  (\bibinfo {year} {2007})}\BibitemShut {NoStop}%
\bibitem [{\citenamefont {Retzker}\ \emph {et~al.}(2005)\citenamefont
  {Retzker}, \citenamefont {Cirac},\ and\ \citenamefont
  {Reznik}}]{Retzker2005}%
  \BibitemOpen
  \bibfield  {author} {\bibinfo {author} {\bibfnamefont {A.}~\bibnamefont
  {Retzker}}, \bibinfo {author} {\bibfnamefont {J.~I.}\ \bibnamefont {Cirac}},
  \ and\ \bibinfo {author} {\bibfnamefont {B.}~\bibnamefont {Reznik}},\ }\href
  {http://link.aps.org/abstract/PRL/v94/e050504} {\bibfield  {journal}
  {\bibinfo  {journal} {Phys. Rev. Lett.}\ }\textbf {\bibinfo {volume} {94}},\
  \bibinfo {eid} {050504} (\bibinfo {year} {2005})}\BibitemShut {NoStop}%
\bibitem [{\citenamefont {Olson}\ and\ \citenamefont
  {Ralph}(2011)}]{RalphOlson1}%
  \BibitemOpen
  \bibfield  {author} {\bibinfo {author} {\bibfnamefont {S.~J.}\ \bibnamefont
  {Olson}}\ and\ \bibinfo {author} {\bibfnamefont {T.~C.}\ \bibnamefont
  {Ralph}},\ }\href {\doibase 10.1103/PhysRevLett.106.110404} {\bibfield
  {journal} {\bibinfo  {journal} {Phys. Rev. Lett.}\ }\textbf {\bibinfo
  {volume} {106}},\ \bibinfo {pages} {110404} (\bibinfo {year}
  {2011})}\BibitemShut {NoStop}%
\bibitem [{\citenamefont {Olson}\ and\ \citenamefont
  {Ralph}(2012)}]{RalphOlson2}%
  \BibitemOpen
  \bibfield  {author} {\bibinfo {author} {\bibfnamefont {S.~J.}\ \bibnamefont
  {Olson}}\ and\ \bibinfo {author} {\bibfnamefont {T.~C.}\ \bibnamefont
  {Ralph}},\ }\href {\doibase 10.1103/PhysRevA.85.012306} {\bibfield  {journal}
  {\bibinfo  {journal} {Phys. Rev. A}\ }\textbf {\bibinfo {volume} {85}},\
  \bibinfo {pages} {012306} (\bibinfo {year} {2012})}\BibitemShut {NoStop}%
\bibitem [{\citenamefont {VerSteeg}\ and\ \citenamefont
  {Menicucci}(2009)}]{Nick}%
  \BibitemOpen
  \bibfield  {author} {\bibinfo {author} {\bibfnamefont {G.}~\bibnamefont
  {VerSteeg}}\ and\ \bibinfo {author} {\bibfnamefont {N.~C.}\ \bibnamefont
  {Menicucci}},\ }\href@noop {} {\bibfield  {journal} {\bibinfo  {journal}
  {Phys. Rev. D}\ }\textbf {\bibinfo {volume} {79}},\ \bibinfo {pages} {044027}
  (\bibinfo {year} {2009})}\BibitemShut {NoStop}%
\bibitem [{\citenamefont {Mart{\'{\i}}n-Mart{\'{\i}}nez}\ and\ \citenamefont
  {Menicucci}(2012)}]{Cosmo}%
  \BibitemOpen
  \bibfield  {author} {\bibinfo {author} {\bibfnamefont {E.}~\bibnamefont
  {Mart{\'{\i}}n-Mart{\'{\i}}nez}}\ and\ \bibinfo {author} {\bibfnamefont
  {N.~C.}\ \bibnamefont {Menicucci}},\ }\href {\doibase
  10.1088/0264-9381/29/22/224003} {\bibfield  {journal} {\bibinfo  {journal}
  {Class. Quantum Gravity}\ }\textbf {\bibinfo {volume} {29}},\ \bibinfo
  {pages} {224003} (\bibinfo {year} {2012})}\BibitemShut {NoStop}%
\bibitem [{\citenamefont {Salton}\ \emph {et~al.}(2015)\citenamefont {Salton},
  \citenamefont {Mann},\ and\ \citenamefont {Menicucci}}]{Salton:2014jaa}%
  \BibitemOpen
  \bibfield  {author} {\bibinfo {author} {\bibfnamefont {G.}~\bibnamefont
  {Salton}}, \bibinfo {author} {\bibfnamefont {R.~B.}\ \bibnamefont {Mann}}, \
  and\ \bibinfo {author} {\bibfnamefont {N.~C.}\ \bibnamefont {Menicucci}},\
  }\href {\doibase 10.1088/1367-2630/17/3/035001} {\bibfield  {journal}
  {\bibinfo  {journal} {New J. Phys.}\ }\textbf {\bibinfo {volume} {17}},\
  \bibinfo {pages} {035001} (\bibinfo {year} {2015})}\BibitemShut {NoStop}%
\bibitem [{\citenamefont {Pozas-Kerstjens}\ and\ \citenamefont
  {Mart\'{i}n-Mart\'{i}nez}(2015)}]{Pozas-Kerstjens:2015}%
  \BibitemOpen
  \bibfield  {author} {\bibinfo {author} {\bibfnamefont {A.}~\bibnamefont
  {Pozas-Kerstjens}}\ and\ \bibinfo {author} {\bibfnamefont {E.}~\bibnamefont
  {Mart\'{i}n-Mart\'{i}nez}},\ }\href {\doibase 10.1103/PhysRevD.92.064042}
  {\bibfield  {journal} {\bibinfo  {journal} {Phys. Rev. D}\ }\textbf {\bibinfo
  {volume} {92}},\ \bibinfo {pages} {064042} (\bibinfo {year}
  {2015})}\BibitemShut {NoStop}%
\bibitem [{\citenamefont {Wald}(1984)}]{Wald1}%
  \BibitemOpen
  \bibfield  {author} {\bibinfo {author} {\bibfnamefont {R.~M.}\ \bibnamefont
  {Wald}},\ }\href@noop {} {\emph {\bibinfo {title} {General Relativity}}}\
  (\bibinfo  {publisher} {The University of Chicago Press},\ \bibinfo {year}
  {1984})\BibitemShut {NoStop}%
\bibitem [{\citenamefont {Mart\'{\i}n-Mart\'{\i}nez}\ \emph
  {et~al.}(2013)\citenamefont {Mart\'{\i}n-Mart\'{\i}nez}, \citenamefont
  {Montero},\ and\ \citenamefont {del Rey}}]{EduardoOld}%
  \BibitemOpen
  \bibfield  {author} {\bibinfo {author} {\bibfnamefont {E.}~\bibnamefont
  {Mart\'{\i}n-Mart\'{\i}nez}}, \bibinfo {author} {\bibfnamefont
  {M.}~\bibnamefont {Montero}}, \ and\ \bibinfo {author} {\bibfnamefont
  {M.}~\bibnamefont {del Rey}},\ }\href {\doibase 10.1103/PhysRevD.87.064038}
  {\bibfield  {journal} {\bibinfo  {journal} {Phys. Rev. D}\ }\textbf {\bibinfo
  {volume} {87}},\ \bibinfo {pages} {064038} (\bibinfo {year}
  {2013})}\BibitemShut {NoStop}%
\bibitem [{\citenamefont {Wald}(1994)}]{Wald2}%
  \BibitemOpen
  \bibfield  {author} {\bibinfo {author} {\bibfnamefont {R.~M.}\ \bibnamefont
  {Wald}},\ }\href@noop {} {\emph {\bibinfo {title} {Quantum Field Theory in
  Curved Spacetime and Black Hole Thermodynamics}}}\ (\bibinfo  {publisher}
  {The University of Chicago Press},\ \bibinfo {year} {1994})\BibitemShut
  {NoStop}%
\bibitem [{\citenamefont {Birrell}\ and\ \citenamefont
  {Davies}(1982)}]{birrelldavies}%
  \BibitemOpen
  \bibfield  {author} {\bibinfo {author} {\bibfnamefont {N.~D.}\ \bibnamefont
  {Birrell}}\ and\ \bibinfo {author} {\bibfnamefont {P.~C.~W.}\ \bibnamefont
  {Davies}},\ }\href {\doibase 10.1017/CBO9780511622632} {\emph {\bibinfo
  {title} {Quantum Fields in Curved Space}}},\ Cambridge Monographs on
  Mathematical Physics\ (\bibinfo  {publisher} {Cambridge University Press},\
  \bibinfo {year} {1982})\BibitemShut {NoStop}%
\bibitem [{\citenamefont {Fewster}\ and\ \citenamefont
  {Verch}(2018)}]{fewster1}%
  \BibitemOpen
  \bibfield  {author} {\bibinfo {author} {\bibfnamefont {C.~J.}\ \bibnamefont
  {Fewster}}\ and\ \bibinfo {author} {\bibfnamefont {R.}~\bibnamefont
  {Verch}},\ }\href@noop {} {\enquote {\bibinfo {title} {Quantum fields and
  local measurements},}\ } (\bibinfo {year} {2018}),\ \Eprint
  {http://arxiv.org/abs/1810.06512} {arXiv:1810.06512 [math-ph]} \BibitemShut
  {NoStop}%
\bibitem [{\citenamefont {Fewster}(2019)}]{fewster2}%
  \BibitemOpen
  \bibfield  {author} {\bibinfo {author} {\bibfnamefont {C.~J.}\ \bibnamefont
  {Fewster}},\ }\href@noop {} {\enquote {\bibinfo {title} {A generally
  covariant measurement scheme for quantum field theory in curved
  spacetimes},}\ } (\bibinfo {year} {2019}),\ \Eprint
  {http://arxiv.org/abs/1904.06944} {arXiv:1904.06944 [gr-qc]} \BibitemShut
  {NoStop}%
\bibitem [{{\relax DLMF}()}]{database}%
  \BibitemOpen
  {\relax DLMF},\ \href {http://dlmf.nist.gov/} {\enquote {\bibinfo {title}
  {{\it NIST Digital Library of Mathematical Functions}},}\ }\bibinfo
  {howpublished} {http://dlmf.nist.gov/, Release 1.0.27 of 2020-06-15},\
  \bibinfo {note} {f.~W.~J. Olver, A.~B. {Olde Daalhuis}, D.~W. Lozier, B.~I.
  Schneider, R.~F. Boisvert, C.~W. Clark, B.~R. Miller, B.~V. Saunders, H.~S.
  Cohl, and M.~A. McClain, eds.}\BibitemShut {Stop}%
\end{thebibliography}%

\end{document}